\documentclass[bibyear]{aa}  
\usepackage{natbib}
\bibpunct{(}{)}{;}{a}{}{,} 

\usepackage{titlesec}

\usepackage{graphicx}
\usepackage{txfonts}
\usepackage{color, xcolor}
\usepackage[colorlinks=true, linkcolor={blue!80!black},citecolor={blue!80!black},urlcolor={magenta}]{hyperref}
\makeatletter
\renewcommand*\aa@pageof{, page \thepage{} of \pageref*{LastPage}}
\makeatother
\begin{document}

   \title{Probing the Temperature Structure of the Inner Region of a Protoplanetary Disk}


   \author{Takahiro Ueda\inst{1,2}
          Satoshi Okuzumi\inst{3}
          Akimasa Kataoka\inst{2}
          \and
          Mario Flock\inst{1}
          }

   \institute{Max-Planck-Institut f{\"u}r Astronomie, K{\"o}nigstuhl 17, D-69117 Heidelberg, Germany\\
              \email{ueda@mpia.de}
             \and
             National Astronomical Observatory of Japan, Osawa 2-21-1, Mitaka, Tokyo 181-8588, Japan
             \and
             Department of Earth and Planetary Sciences, Tokyo Institute of Technology, Meguro, Tokyo, 152-8551, Japan
             }

   \date{Received XXX; accepted YYY}

 
  \abstract
   {Disk temperature structure is crucial for the formation of planets. Midplane heating induced by disk accretion plays a key role in determining the disk temperature particularly at the inner disk midplane where planets form. However, the efficiency of accretion heating has been not well constrained by observations.}
   {Our aim is to observationally constrain the physical properties of the inner region of the CW Tau disk where the midplane heating potentially takes place.}
   {We construct two-dimensional physical models of the CW Tau disk, taking into account the midplane heating.
   The models are compared with the ALMA dust continuum observations at Bands 4, 6, 7 and 8, with an angular resolution of 0\farcs1. 
   The observed brightness temperatures are almost wavelength-indenpendent at $\lesssim$10 au.
   }
   {We find that if the maximum dust size $a_{\rm max}$ is $\lesssim100~{\rm \mu m}$, the brightness temperatures predicted by the model exceed the observed values, regardless of the efficiency of accretion heating.
    The low observed brightness temperatures can be explained if millimeter scattering reduces the intensity.
    If the disk is passive, $a_{\rm max}$ needs to be either $\sim150~{\rm \mu m}$ or $\gtrsim$ few ${\rm cm}$.
    The accretion heating significantly increases the brightness temperature particularly when $a_{\rm max}\lesssim300~{\rm \mu m}$, and hence $a_{\rm max}$ needs to be either $\sim300~{\rm \mu m}$ or $\gtrsim$ few ${\rm cm}$.
    The midplane temperature is expected to be $\sim$1.5--3 times higher than the observed brightness temperatures, depending on the models.
    The dust settling effectively increases the temperature of the dust responsible for the millimeter emission in the active disk, which makes the model with $300~{\rm \mu m}$-sized dust overpredicts the brightness temperatures when strong turbulence is absent.
    Porous dust (porosity of 0.9) makes the accretion heating more efficient so that some sort of reduction in accretion heating is required.
   }
   {The brightness temperature is not a simple function of the dust temperature because of the effect of scattering and midplane heating even if the disk is optically thick.
   The current data of the CW Tau disk are not enough to discriminate between the passive and active disk models.
   Future longer wavelength and higher angular resolution observations will help us constrain the heating mechanisms of the inner protoplanetary disks.
   }


   \authorrunning{Ueda et al.}
   \maketitle
%

\section{Introduction}

The temperature structure of inner protoplanetary disks controls the evolution of dust grains (e.g., \citealt{Birnstiel+10,Okuzumi+16,DA17,Ueda+21b}), subsequent formation of planets and their migration (e.g., \citealt{KN12,Bitsch+14,Bitsch19,SB21}) and eventually the chemical composition of planets (e.g., \citealt{Oberg+11,Madhusudhan+14,OU21,SchneiderBitsch21,Notsu+22}).
Particularly at the midplane of the inner regions of disks ($\lesssim10$ au), the heating resulting from disk gas accretion is thought to play a key role in determining the disk temperature.
In the classical model of accretion heating, the gravitational energy of the disk gas is released near the midplane where the energy is hard to escape from the disk, leading to efficient heating of the disk midplane (e.g., \citealt{Hubeny90,NN94}).
However, recent non-ideal magneto-hydrodynamical models have shown that accretion heating may be less efficient compared to the classical model when the accretion is primarily driven by the magnetorotational instability (MRI) in the upper layer of the disk \citep{HT11} or by magneto-hydrodynamical disk winds \citep{Mori+19}.
This is because the gravitational energy released at the upper layer can escape from the disk more easily.
Still, accretion heating in magneto-hydrodynamically accreting disks can affect the disk temperature structure depending on the disk ionization state and opacity \citep{BL20,Kondo+22}.

Even though how the disk midplane is heated is crucial for the formation of planets, it is poorly constrained by observations.
The Atacama Large Millimeter/submillimeter Array (ALMA) provides an opportunity to probe the temperature structure of the inner regions of disks, where accretion heating potentially takes place.
The ALMA high-resolution multi-wavelength analysis allows us to evaluate the dust properties (e.g., dust size, temperature and surface density) as a function of radial distance based on the spectral behavior of the (sub-)millimeter dust thermal emission \citep{CG+19,Macias+21,Sierra+21,Ueda+22,Guidi+22}.
However, previous studies have generally assumed that the observed dust temperatures at different ALMA wavelengths are identical (i.e., vertically isothermal). 
If the disk is optically thick at the observing wavelengths, the different observing wavelengths trace different heights within the disk, which may alter the interpretation of the disk properties (\citealt{SL20,Ueda+21}).

In other words, the vertical disk structure can be probed by the multi-wavelength sub-millimeter to centimeter observations by leveraging the difference in optical depth at each wavelength.
For instance, recent ALMA observations of CO isotopologue line emissions have provided insights into the vertical structure of the outer regions of disks by utilizing the expected emission heights of different CO isotopologues (\citealt{Law+21,Law+23}).
For the inner regions of disks, dust continuum observations would be more suitable than the gas molecule observations as it is easier to achieve better angular resolution and sensitivity with a reasonable observing time.
\citet{Okuzumi+23} demonstrated that the vertical temperature structure of the inner few au of disks can be inferred from the multi-wavelength dust continuum observations using ALMA and the next generation Very Large Array (ngVLA).

In this paper, we investigate the temperature structure of the disk around CW Tau using ALMA multi-wavelength dust continuum observations. 
The CW Tau disk has a high accretion rate ($3$--$10\times10^{-8}M_{\odot}~{\rm yr^{-1}}$; \citealt{McClure+19,Robinson+22,Gangi+22}) and has been observed at ALMA Bands 4, 6, 7 and 8 \citep{Ueda+22}.
This combination of a high accretion rate and rich observational data makes the CW Tau disk an excellent target for studying the vertical temperature structure of the inner region potentially heated by disk accretion.
This paper is constructed as follows.
In Section \ref{sec:method}, we introduce our observational data and theoretical models.
The models are compared with the observations in Section \ref{sec:comparison}.
Discussion and summary are in Section \ref{sec:discussion} and \ref{sec:summary}, respectively.

\section{Methods} \label{sec:method}

\subsection{ALMA observations} \label{sec:obs}
We make use of the ALMA data of the CW Tau disk taken and calibrated in \citet{Ueda+22}.
The observations were carried out at ALMA Bands 4 ($\lambda=2.17~{\rm mm}$), 6 ($1.34~{\rm mm}$), 7 ($0.89~{\rm mm}$) and 8 ($0.75~{\rm mm}$).
The CW Tau disk is a Class II disk with the stellar luminosity of $L_{*}=0.45L_{\odot}$ \citep{HH14}, accretion luminosity of $L_{\rm acc}=0.85L_{\odot}$ \citep{Gangi+22} and accretion rate of $\dot{M}=3$--$10\times10^{-8}M_{\odot}~{\rm yr^{-1}}$ (\citealt{McClure+19,Robinson+22,Gangi+22}) which is  much higher than the typical value of the Class II disks (e.g., $\sim3.6\times10^{-9}M_{\odot}~{\rm yr}^{-1}$; \citealt{Manara+17} see also \citealt{Manara+22}).
Its high accretion rate and relatively low stellar luminosity makes the CW Tau disk suitable for studying accretion heating of the disk.
In this work, we adopt $\dot{M}=4\times10^{-8}M_{\odot}~{\rm yr}^{-1}$, which falls within the range of observed values for CW Tau.

\begin{figure}[ht]
\begin{center}
\includegraphics[scale=0.47]{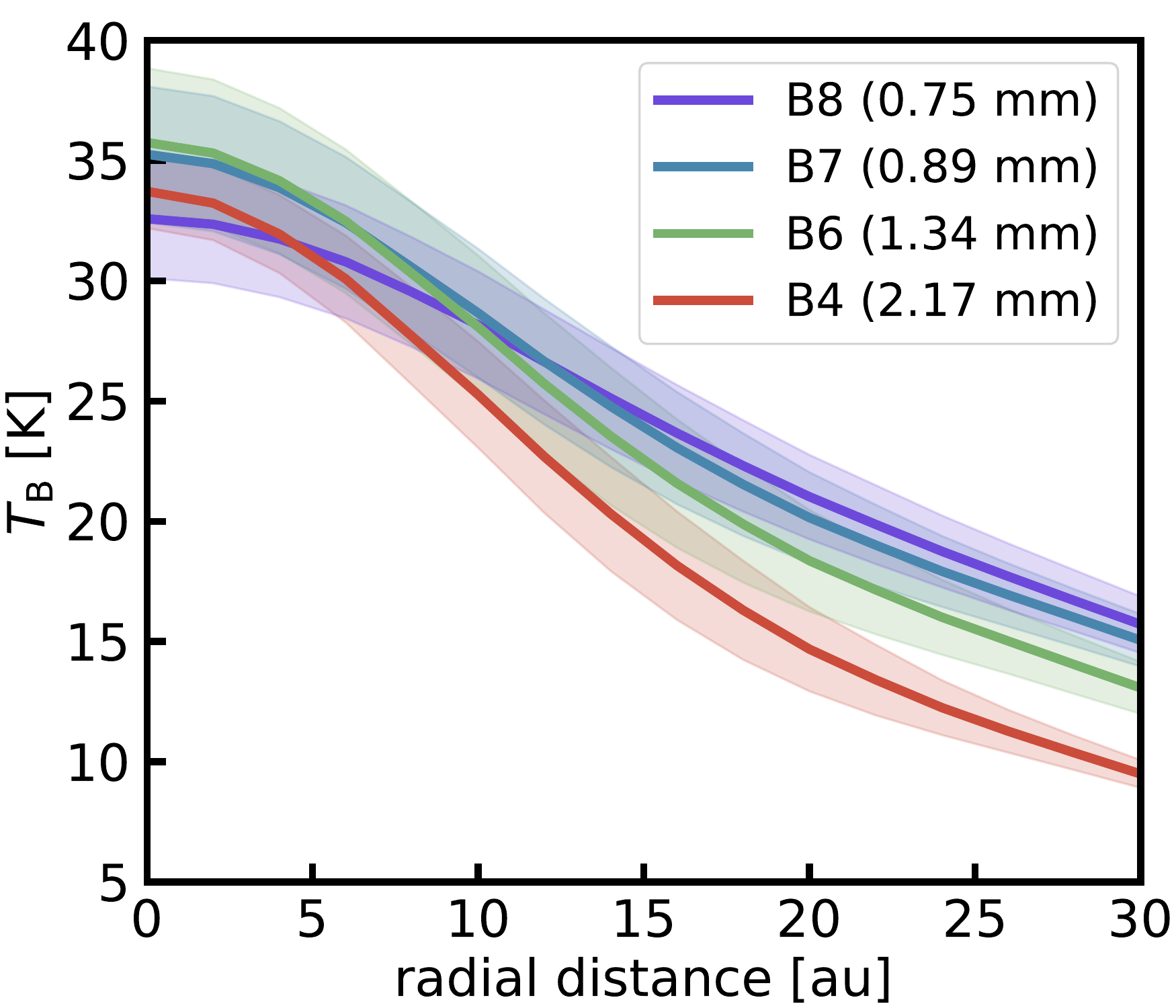}
\caption{
Brightness temperature profile observed at ALMA Bands 4, 6, 7 and 8 obtained by \citet{Ueda+22}.
The transparent region denotes the uncertainty arising from both flux calibration and thermal noise.
}
\label{fig:tb_obs}
\end{center}
\end{figure}

Figure \ref{fig:tb_obs} presents the observed brightness temperature at ALMA Bands 4, 6, 7 and 8.
The observational data has a common angular resolution of 0$\farcs$1, corresponding to a spatial resolution of 13.2 au for the distance of CW Tau (132 pc; \citealt{Gaia+21}).
The brightness temperature at ALMA Bands 4, 6, 7, and 8 shows a nearly identical value of $\sim35$ K at $\lesssim10$ au.
At the outer region, $\gtrsim20$ au, the brightness temperature exhibits variations among the different ALMA bands. 
In particular, the brightness temperature at Band 4 is lower compared to the other bands.

\subsection{Theoretical model} \label{sec:models}
We construct two-dimensional (radial position $r$ and vertical height $z$) models of the CW Tau disk in order to compare them with the ALMA observations. In this section, we provide a detailed description of our theoretical models of the disk structure.

\subsubsection{Temperature structure}\label{sec:midplane}
The temperature structure of the disk is determined by the combined effects of stellar irradiation and accretion heating:
\begin{eqnarray}
T^{4}=T_{\rm irr}^{4}+T_{\rm acc}^{4},
\end{eqnarray}
where $T_{\rm irr}$ and $T_{\rm acc}$ denote the disk temperature determined by stellar irradiation and accretion heating, respectively.
The temperature structure of the passively irradiated disk is calculated as 
\begin{eqnarray}
T_{\rm irr}^{4} = \frac{L_{*}+L_{\rm acc}}{16\pi \sigma r^{2}} \left\{\exp{\left(-\tau_{*}\right)} + 2\psi\right\}, \label{eq:t_irr}
\end{eqnarray}
where $\sigma$ is the Stefan-Boltzmann constant, $\tau_{*}$ is the radial optical depth for stellar irradiation.
The angle $\psi$ represents the angle between the ray from the central star and the disk surface. 
Its actual value depends on the distance from the central star and the detailed disk surface structure.
We assume $\psi=0.02$ which corresponds to $h_{*}/r\sim0.07$ with $h_{*}$ being the height of the disk photosphere above the midplane (\citealt{DDC01}).

Equation \eqref{eq:t_irr} provides the black body temperature in the optically thin regime when $\tau_{*}\ll1$, whereas it yields temperature profile of the classical optically thick passive disk in the limit of $\tau_{*}\gg1$ (e.g., \citealt{DDC01}; see also \citealt{Huang+18}).
In the optically thick interior region ($\tau_{*}\gg1$), the temperature structure is vertically isothermal.
It is worth noting that the temperature structure of the optically thin region $(\tau_{*}\ll1)$ has no significant impact on our analysis as the emission observed by ALMA is dominated by the emission from the region where $\tau_{*}\gg1$.
The temperature of optically thick region of the passively-heated disk can be lower than our model because of scattering of the stellar irradiation (\citealt{OUT22}), which is not included in our model.
With our dust model, this could reduce the disk temperature by 10\%. 
We examine the validity of Equation \eqref{eq:t_irr} in Appendix \ref{sec:temp}.

The temperature structure determined by accretion heating is (e.g., \citealt{NN94,SL20})
\begin{eqnarray}
T_{\rm acc}^{4} = \frac{3}{4}\left(\tau_{\rm z}+\frac{2}{3}\right) \frac{3\dot{M}\Omega_{\rm K}^{2}}{8\pi\sigma},
\end{eqnarray}
where $\Omega_{\rm K}$ is the Keplerian frequency and $\tau_{\rm z}$ is the vertical optical depth for the dust thermal emission integrated from each position to $z=\infty$ (or $z=-\infty$ if $z<0$);
\begin{eqnarray}
\tau_{\rm z}(z) = \int_{z}^{\infty}\kappa_{\rm R}(z^{\prime})\rho_{\rm d}(z^{\prime}) dz^{\prime},
\end{eqnarray}
with $\kappa_{\rm R}$ and $\rho_{\rm d}$ being the Rosseland-mean opacity and dust density, respectively. 
Because $\tau_{\rm z}$ increases as $z$ approaches the midplane, $T_{\rm acc}$ increases as $z$ approaches the midplane. 
We note that the actual value of $\dot{M}$ can be lower/higher than our adopted value by a factor of $\sim 2$ (see Section \ref{sec:obs}), which decrease/increase $T_{\rm acc}$ by $\sim$20\%.
As $T_{\rm acc}$ is a function of the product of $\tau_{\rm z}$ and $\dot{M}$, the disk temperature structure is almost identical as long as $\tau_{\rm z}\dot{M}$ is identical.

In this study, we consider two scenarios: a passive disk where the temperature structure is solely determined by stellar irradiation (i.e., $T^{4}=T_{\rm irr}^{4}$), and an active disk where additional heating due to accretion is taken into account ($T^{4}=T_{\rm irr}^{4}+T_{\rm acc}^{4}$). 
As shown in Section \ref{sec:comparison}, accretion heating dominates over stellar irradiation ($T\approx T_{\rm acc}$) near the midplane within the inner few to few dozen au, depending on the specific disk models.

\subsubsection{Dust opacities}\label{sec:opac}
The dust opacity plays a crucial role in determining both the temperature structure of the disk and the (sub-)millimeter emission observed with ALMA.
The dust composition is assumed to be that of the DSHARP model \citep{Birnstiel+18} which is a mixture of water ice \citep{WB08}, astronomical silicates \citep{Draine03}, troilite and refractory organics \citep{HS96}.
It is important to note that the dust composition remains highly uncertain, and therefore, we also present modeling results using an alternative dust model, the DIANA dust model \citep{Woitke+16} in Appendix \ref{sec:comp}.

The dust size distribution ranges from $a_{\rm min}$ to $a_{\rm max}$ with a power-law index of $-p_{\rm d}$.
We assume $p_{\rm d}=3.5$ which is similar to the so-called MRN distibution \citep{MRN} as a fiducial case, while more flat profile ($p_{\rm d}=2.5$) is also applied to investigate its impact on the brightness temperature.
The latter corresponds to more top-heavy distribution where the dust opacity is dominated by larger grains.
The minimum dust size $a_{\rm min}$ is set to be $0.05~{\rm \mu m}$.
We confirmed that our conclusions remain unchanged even if we adopt larger $a_{\rm min}$, $0.5~{\rm \mu m}$.
We consider two different dust porosity values, namely $p=0$ (corresponding to compact dust) and $0.9$ (corresponding to a filling factor of $f=1$ and $0.1$, respectively). 
We do not consider very porous dust ($p\gg0.9$) because extremely porous dust may not account for the high polarization degree of the dust thermal emission observed toward the CW Tau disk (\citealt{Bacciotti+18,Tazaki+19,Kirchschlager+19}).
The opacities are computed with Optool \citep{optool}.

\begin{figure}[t]
\begin{center}
\includegraphics[scale=0.44]{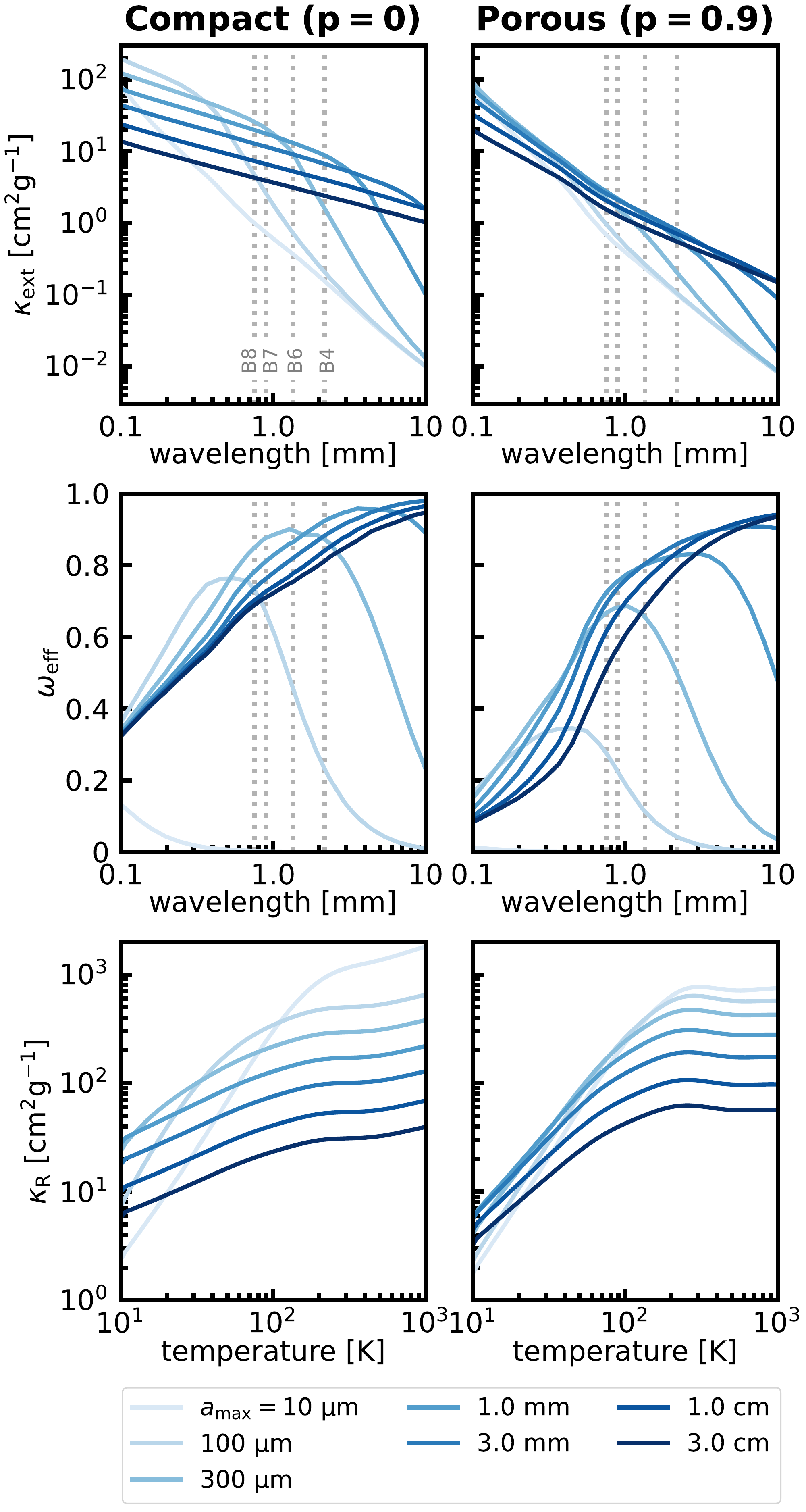}
\caption{Effective extinction opacity (top), effective albedo (middle) and Rosseland-mean opacity (bottom) of compact ($p=0$; left) and porous ($p=0.9$; right) dust with different $a_{\rm max}$.
The power-law index of the dust size distribution is set to $p_{\rm d}=3.5$.
The vertical gray dotted lines denote the wavelength of ALMA Bands 4, 6, 7 and 8.
}
\label{fig:opac}
\end{center}
\end{figure}

Figure \ref{fig:opac} shows the effective extinction opacity $\kappa_{\rm ext}$, effective albedo $\omega_{\rm eff}$ and Rosseland-mean opacity $\kappa_{\rm R}$ of our dust model. 
The effective extinction opacity is defined as $\kappa_{\rm ext}=\kappa_{\rm abs}+\kappa_{\rm sca}(1-g)$, where $\kappa_{\rm abs}, \kappa_{\rm sca}$ and $g$ are the absorption opacity, scattering opacity and anisotropic scattering parameter, respectively.
The effective albedo is given as $\omega_{\rm eff}=\kappa_{\rm sca}(1-g)/\kappa_{\rm ext}$.
The extinction opacity has a slope of $\kappa_{\rm ext}\propto\lambda^{-1.7}$ when $a_{\rm max}\ll\lambda/2\pi$, whereas it is flatter when $a_{\rm max}\gg\lambda/2\pi$.
In the intermediate regime ($a_{\rm max}\sim\lambda/2\pi$), the opacity slope is steeper than $-1.7$ because of the Mie interference.
The interference is significant only for compact grains, leading to a higher opacity compared to porous grains \citep{Kataoka+14}.

The effective scattering albedo at the ALMA wavelengths ($\sim$0.8--2 mm) is $\lesssim0.2$ for $a_{\rm max}\ll\lambda/2\pi\sim300~{\rm \mu m}$. 
However, for $a_{\rm max}\gtrsim300~{\rm \mu m}$, the scattering albedo reaches 0.8 or even higher at the ALMA wavelengths.
The high scattering albedo efficiently reduces the emergent intensity at the ALMA wavelengths, which makes the disk fainter than that without scattering (e.g., \citealt{Liu19,Zhu+19}).
At the ALMA wavelengths, compact dust has slightly higher $\omega_{\rm eff}$ compared to porous dust.
This is because porous dust scatters incident radiation more efficiently in the forward direction, resulting in a lower effective albedo.
We note that the absorption and scattering opacity can be scaled by $a_{\rm max}f$ except for the behavior of the Mie interference \citep{Kataoka+14}. 
However, the asymmetry parameter $g$ depends on the porosity, which makes the difference in the optical properties of compact and porous dust for a given $a_{\rm max}f$ \citep{Zhang+23}.

\subsubsection{Dust disk}
The dust surface density is given by
\begin{eqnarray}
\Sigma_{\rm d}=\frac{\tau_{\rm 10,B4}}{\kappa_{\rm ext,B4}}\left(\frac{r}{\rm 10~au}\right)^{-1}, 
\end{eqnarray}
where $\tau_{\rm 10,B4}$ is the vertical extinction optical depth at ALMA Band 4 at 10 au.
Although the true dust surface density is highly uncertain, the wavelength-independent brightness temperature implies that the inner disk region is at least moderately optically thick at these wavelengths (see \citealt{Ueda+22}).
In our model, we adopt $\tau_{\rm 10,B4}=3$ for all models.
In Appendix \ref{sec:app1}, we show the effect of $\tau_{\rm 10,B4}$ on the brightness temperature.
We confirmed that $\tau_{\rm 10,B4}$ needs to be larger than unity in order to match the observed brightness temperature at ALMA Band 4. 
We will discuss the estimated dust and gas surface densities as well as the corresponding accretion efficiency parameter $\alpha_{\rm acc}$ in Section \ref{sec:sigmad}.

In the vertical direction, the dust density distribution is given by:
\begin{eqnarray}
\rho_{\rm d}=\frac{\Sigma_{\rm d}}{\sqrt{2\pi}h_{\rm d}} \exp{\left(-\frac{z^{2}}{2h_{\rm d}^{2}} \right)},
\end{eqnarray}
where $h_{\rm d}$ represents the scale height of the dust disk.
In our fiducial model, we assume that the dust is well-coupled with the gas, i.e., $h_{\rm d}=h_{\rm g}$ with $h_{\rm g}$ being the gas scale height.
The gas scale height is calculated by $h_{\rm g}=c_{\rm s}/\Omega_{K}$ with $c_{\rm s}$ being the sound speed of the gas at the midplane.
We also investigate the effect of dust settling.
In incorporate dust settling, we define the Stokes number as
\begin{eqnarray}
{\rm St} = \frac{\pi}{2}\frac{\rho_{\rm m}fa}{\Sigma_{\rm g}},
\end{eqnarray}
where $\rho_{\rm m}=1.675~{\rm g~cm^{-3}}$ is the material density, $f=1-p$ is the dust filling factor and $\Sigma_{\rm g}$ is the gas surface density and $a$ is the dust radius.
For simplicity, we assume that the gas surface density is 100 times larger than the dust surface density.
The dust scale height is assumed to be in mixing-settling equilibrium \citep{Dubrulle+95,YL07};
\begin{eqnarray}
h_{\rm d} = \left(1+\frac{\rm St}{\alpha_{\rm t}}\frac{1+2{\rm St}}{1+{\rm St}}\right)^{-1/2}h_{\rm g},
\end{eqnarray}
where $\alpha_{\rm t}$ is the turbulence strength for vertical dust mixing.
In our model with settling, we assume relatively weak vertical mixing with $\alpha_{\rm t}=10^{-4}$ in order to investigate the difference from the no-settling limit.

The computational domain extends radially from $r=0.5$ to 50 au, logarithmically divided into 100 cells. 
For the polar direction, the calculation domain covers from $\theta=-\pi/6$ to $\pi/6$ (where $\theta=0$ gives disk midplane) with 400 uniformly spaced cells.

To investigate the impact of different physical mechanisms on the brightness temperature, we constructed models with various setups. 
The details of these models are summarized in Table \ref{table:models}.

\begin{table}[ht]
  \begin{center}
  \caption{Summary of our models}\label{table:models}
  \begin{tabular}{c|c|c|c|c}
 Model name & Accretion & Dust     & $p$ & $p_{\rm d}$ \\
            & heating   & settling &     &  \\
  \hline
 Passive disk & off & off & 0 & 3.5\\
 Fiducial active disk & on & off & 0 & 3.5\\
 Settling & on & on & 0 & 3.5\\
 Top-heavy & on & off & 0 & 2.5\\
 Porous & on & off & 0.9 & 3.5\\
\hline
  \end{tabular}
  \end{center}
\end{table}

\begin{figure*}[t]
\begin{center}
\includegraphics[scale=0.45]{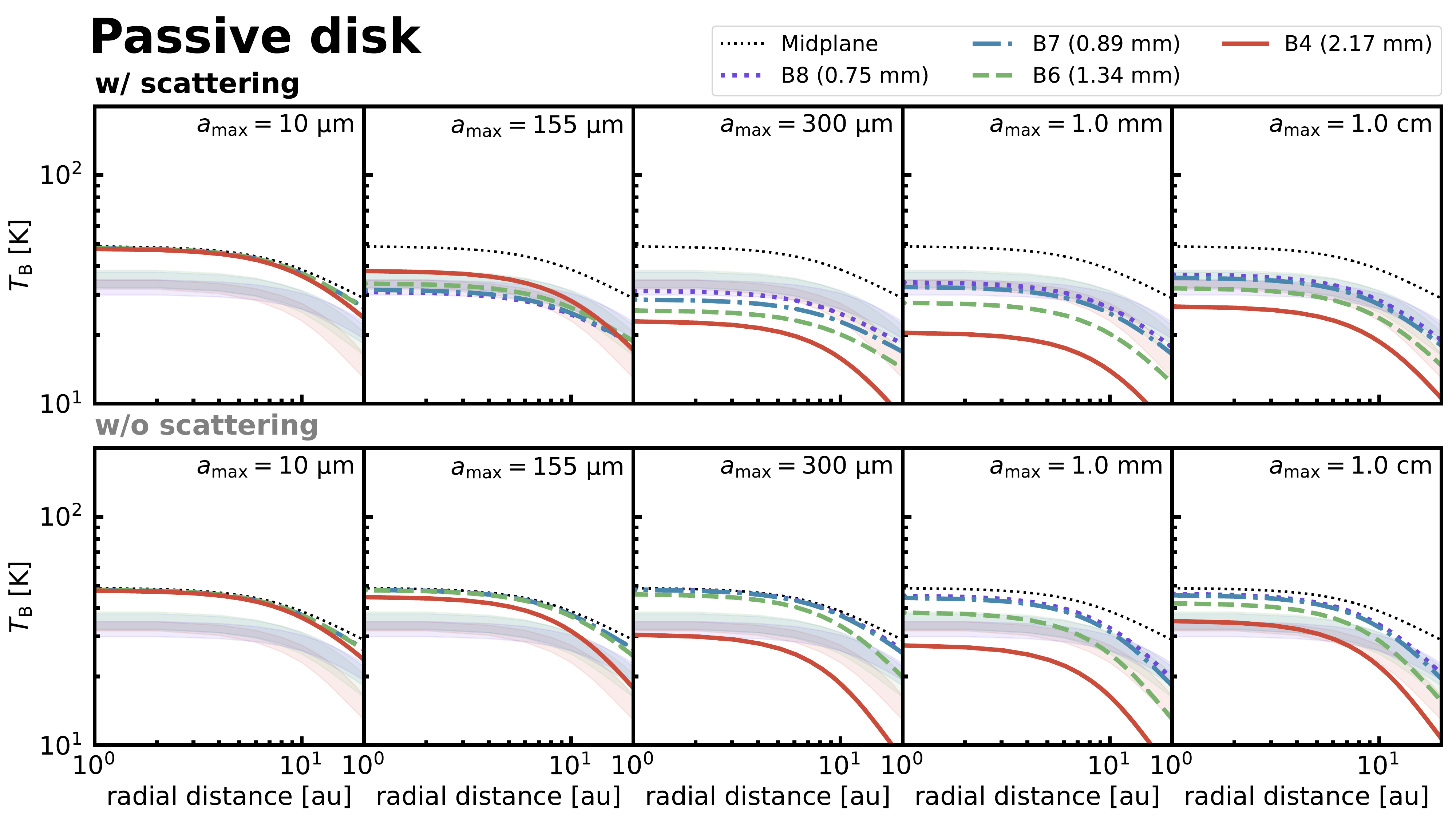}
\caption{
Brightness temperature profile of the passive disk model ($p=0$, $\tau_{\rm 10,B4}=3$ and $p_{\rm d}=3.5$) with (top) and without (bottom) scattering with different $a_{\rm max}$.
The transparent shaded region shows the observed brightness temperature.
The model brightness temperature is calculated after the intensity is convolved with the beam size of 0$\farcs$1 which corresponds to the physical distance of 13.2 au.
The black dotted line denote the midplane temperature profile convolved with a gaussian with a full width at half maximum (FWHM) of 13.2 au.
}
\label{fig:1dtemp_noacc}
\end{center}
\end{figure*}

\subsection{Imaging simulation}
Given the disk structure obtained in Section \ref{sec:models}, we perform Monte Carlo radiative transfer calculations using RADMC-3D \citep{RADMC} to generate synthetic images.
The dust opacities are the same as those described in Section \ref{sec:opac}.
We use $10^{7}$ photon packages for each simulation.
To compare the synthetic images with the observations, we smooth the images with the observing beam size of $0\farcs1$, which corresponds to a physical distance of 13.2 au for CW Tau ($d=132$ pc). 
The disk inclination is assumed to be $58^{\circ}$ \citep{Ueda+22} and the intensity is averaged over the azimuthal direction to obtain the radial intensity profile.
The radial intensity profile is then converted into the brightness temperature using the full Planck function.
To account for anisotropic scattering, we employ the Henyey-Greenstein approximation \citep{HG41}.

\section{Comparison of models and observations} \label{sec:comparison}

In this section, we compare our theoretical models with the observations.

\subsection{Passive disk models}
Let us start the comparison from the simplest cases: passive disks, where the temperature structure is determined solely by stellar irradiation.

Figure \ref{fig:1dtemp_noacc} shows the brightness temperature profile of the passive disk model for different $a_{\rm max}$.
In Figure \ref{fig:1dtemp_noacc}, we show the brightness temperature computed with and without scattering.
When scattering is considered, the emergent intensity is reduced, leading to lower brightness temperatures compared to the actual dust temperature, even in optically thick disks (\citealt{Liu19,Zhu+19,Ueda+20}).

For $a_{\rm max}=10~{\rm \mu m}$, the model brightness temperatures at $r\lesssim10$ au are identical for all wavelengths and are $\sim1.5$ times higher than observed.
Scattering has little effects on the brightness temperatures in this case because of the low albedo of the $10~{\rm \mu m}$-sized grains (see the center panel of Figure \ref{fig:opac}).
At $\gtrsim15$ au, the brightness temperature at Band 4 is slightly lower than those at the other bands because that region is marginally optically thin at wavelengths $\gtrsim$ 2 mm.

For $a_{\rm max}\gtrsim100~{\rm \mu m}$, the brightness temperatures are comparable to or lower than those for $a_{\rm max}=10~{\rm \mu m}$.
The observed brightness temperature in our models without scattering is affected by the optical depth of the disk. In our model, the dust surface density is adjusted such that the extinction optical depth at ALMA Band 4 reaches a value of 3 at a radial distance of 10 au ($\tau_{\rm 10,B4}=3$).
In the models without scattering, the effect of scattering is taken into account in the calculation of the dust surface density. This means that the dust surface density remains the same for models with and without scattering. However, when calculating the emergent intensity, scattering is ignored.
Because of this setup, the vertical optical depth is effectively smaller for the models without scattering if scattering dominates over absorption.
For instance, dust with $a_{\rm max}=300~{\rm \mu m}$ has the effective albedo of $\sim0.9$ at ALMA Band 4, which corresponds to the absorption optical depth of $\tau_{\rm abs,10,B4}=\tau_{\rm ext,10,B4}(1-\omega_{\rm eff})\sim0.3$. 
Therefore, the no-scattering model with $a_{\rm max}=300~{\rm \mu m}$ is optically thin at ALMA Band 4 and hence shows lower brightness temperature.
For the models with scattering, the model disks are always optically thick at $<10$ au at all wavelengths we considered.

We see from Figure \ref{fig:1dtemp_noacc} that the brightness temperatures from the passive disk models without scattering exceed those from the observations.
This implies that the scattering-induced intensity reduction takes place in the inner region of the CW Tau disk.
For $a_{\rm max}\sim150~{\rm \mu m}$, the brightness temperatures are reduced by $\sim$30\% by scattering, which makes the passive disk with $150~{\rm \mu m}$-sized dust consistent with the observations.
The model with $a_{\rm max}=300~{\rm \mu m}$ has a high effective albedo ($\gtrsim0.8$) at all wavelengths considered, yielding brightness temperatures lower than observed.
For $a_{\rm max}\gtrsim300~{\rm \mu m}$, the effective albedo at ALMA Bands 7-8 decreases as dust size increases, which makes the models with large grains are consistent with the observations at ALMA Bands 7-8.
However, at Band 4, dust with sizes in the range of $300~{\rm \mu m}\lesssim a_{\rm max}\lesssim1~{\rm mm}$ still exhibit a high albedo, leading to brightness temperatures that are too low compared to the observed values. 
Because the scattering albedo at ALMA Band 4 decreases for $a_{\rm max}\gtrsim1~{\rm mm}$, very large dust with $a_{\rm max}\gtrsim10~{\rm cm}$ may account for the observations.

\begin{figure}[ht]
\begin{center}
\includegraphics[scale=0.53]{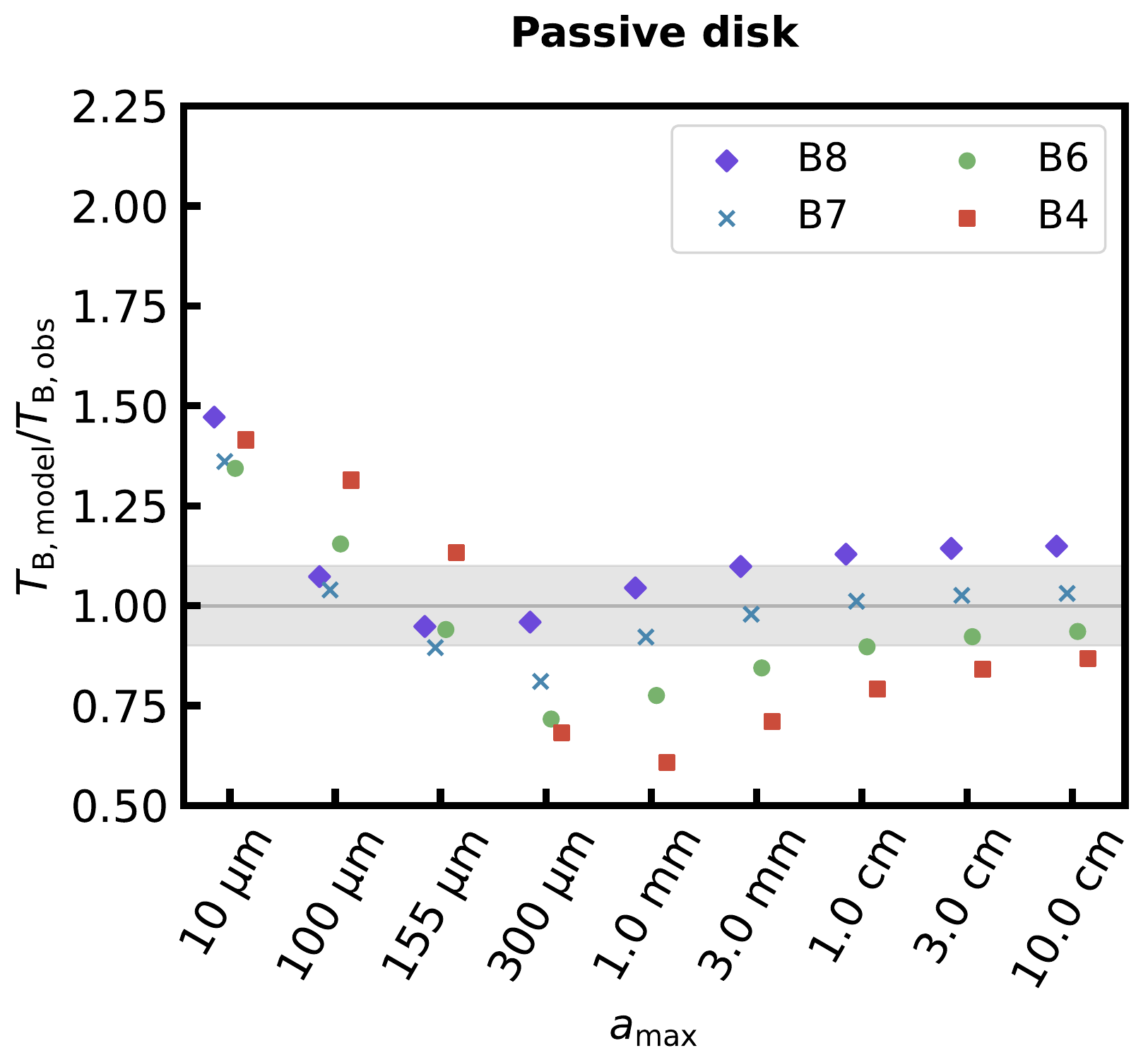}
\caption{
Brightness temperature at the center of model images normalized by the observed brightness temperature.
The gray shaded region denotes the region where the model explains the observations with an accuracy of 10\%, which is the typical ALMA calibration uncertainty.
}
\label{fig:passive_center}
\end{center}
\end{figure}

Figure \ref{fig:passive_center} presents the brightness temperature at the center of the model images normalized by the observed value, providing insight into the temperature structure within the central region with diameter of $\sim13$ au, which corresponds to the size of one beam.
We see that the brightness temperature of models with $a_{\rm max}\lesssim100~{\rm \mu m}$ exceeds the observed values by $\sim30\%$.
In contrast, the brightness temperatures are significantly lower than the observations if $a_{\rm max}=300~{\rm \mu m}$.
The intermediate size, $a_{\rm max}\sim150~{\rm \mu m}$, reasonably explains the observations.
If dust size is larger than millimeter, the model brightness temperatures are more similar to the observed values as dust size increases, showing that very large grains ($\gtrsim$ few cm) are roughly consistent with the observations.

In summary, when considering a disk heated solely by stellar irradiation, two possible ranges for the maximum dust radius emerge to explain the observed brightness temperatures: $\sim150~{\rm \mu m}$ or $\gtrsim$ few cm.
Very small dust grains, $a_{\rm max}\ll150~{\rm \mu m}$, are unlikely to account for the observations due to their low effective albedo at millimeter wavelengths, which results in higher brightness temperatures than observed. 
On the other hand, millimeter-sized dust grains are also improbable as their high effective albedo at ALMA Band 4 leads to significantly lower model brightness temperatures compared to the observed values.

\subsection{Active disk models}
The passive disk models considered in the previous subsection ignore any internal heating sources. 
However, the inner region of the disk can be heated by gas accretion, and hence, the disk may have a vertical temperature gradient that affects the brightness temperature. In this subsection, we will show how accretion heating affects the brightness temperature of the CW Tau disk.

\subsubsection{Case of $a_{\rm max}=10~{\rm \mu m}$}\label{sec:10um}
To clarify the impact of the vertical temperature structure on the ALMA brightness temperatures, we start the analysis by considering the scenario of $a_{\rm max}=10~{\rm \mu m}$, wherein scattering of dust thermal emission is negligible.
The most notable factor in this case is the substantial vertical optical depth $\tau_{\rm z}$, which establishes the accretion heating as the primary determinant of the inner disk temperature.

\begin{figure}[ht]
\begin{center}
\includegraphics[scale=0.59]{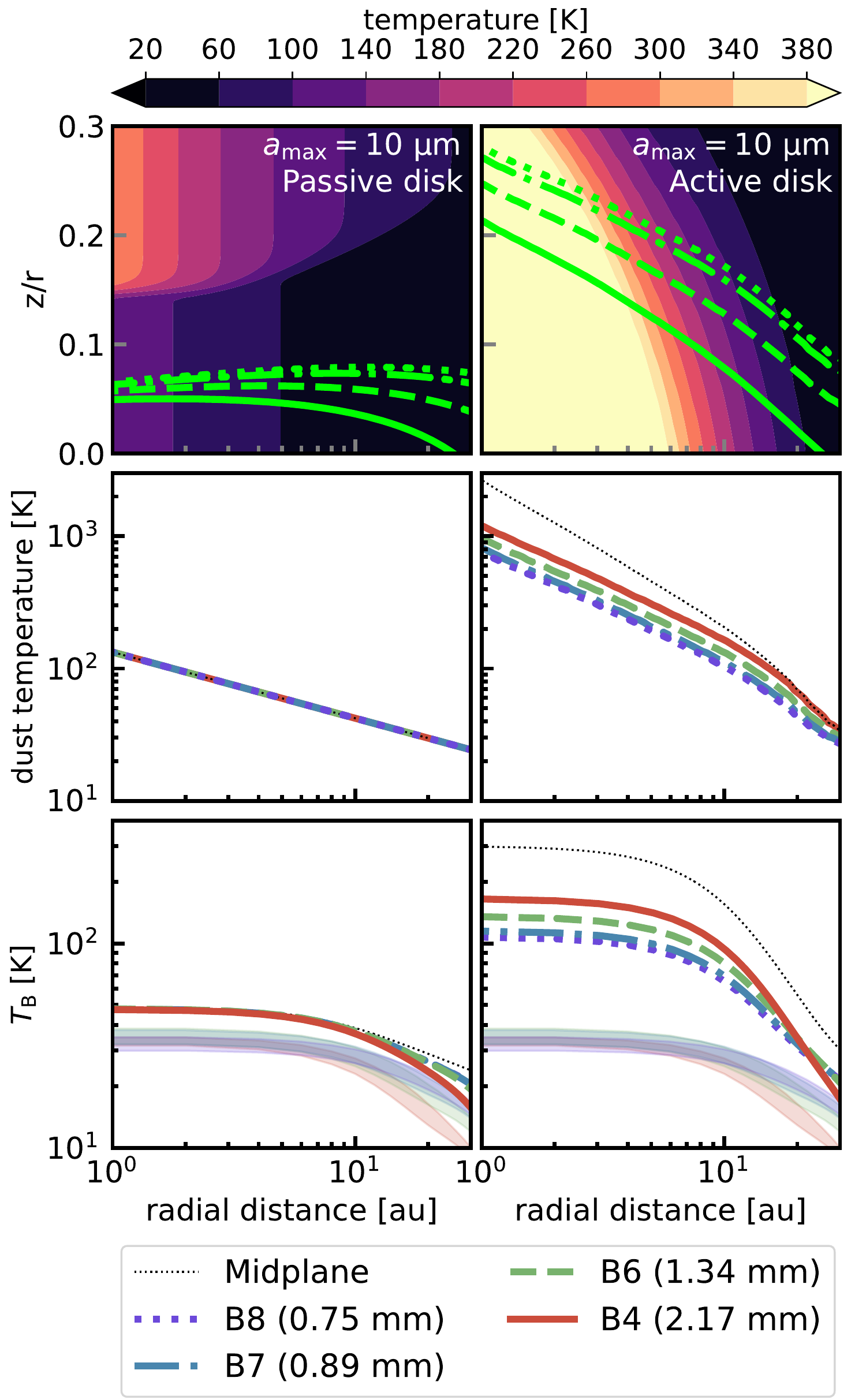}
\caption{
Two-dimensional temperature structure (top), dust temperature at the ALMA surfaces (middle) and synthetic brightness temperature profiles (bottom).
The maximum dust size is $10~{\rm \mu m}$. 
The left panels show the passive disk model where the disk is heated only by the stellar irradiation, whereas the right panels show the active disk model where the disk is heated by both stellar irradiation and disk accretion.
The light green lines in the top panels denote the height where the vertical extinction optical depth reaches unity (ALMA surfaces) at ALMA Bands 4 (solid), 6 (dashed), 7 (dash-dotted) and 8 (dotted).
The dust temperatures shown in the middle panels are not beam-convolded, while the brightness temperatures shown in the bottom panels are taken from the synthetic images convolved with a gaussian with a FWHM of 13.2 au.
The transparent regions denote the observed brightness temperatures.
}
\label{fig:2dtemp_10um}
\end{center}
\end{figure}

Figure \ref{fig:2dtemp_10um} shows the two-dimensional temperature structure of the disk with $a_{\rm max}=10~{\rm \mu m}$, both with and without considering the effects of accretion heating.
It is evident that within a radius of approximately 20 au, the temperature structure is primarily influenced by accretion heating when $a_{\rm max}=10~{\rm \mu m}$.
Additionally, Figure \ref{fig:2dtemp_10um} presents the dust temperatures at the surfaces where the extinction optical depths at ALMA wavelengths reach unity (hereafter ALMA surfaces).
It is worth noting that despite having identical total vertical optical depths in the passive and active disk models, the ALMA surfaces are higher in the active disk model due to the elevated gas scale height compared to the passive model.

The dust temperature at the ALMA surfaces remains constant regardless of the observing wavelengths in the case of a passively heated disk. However, in the presence of accretion heating, the dust temperature at the ALMA wavelengths becomes wavelength-dependent.
This dependence arises because longer observing wavelengths probe regions closer to the midplane, where accretion heating is more efficient. Consequently, longer observing wavelengths result in higher dust temperatures. 
For instance, at 10 au, the dust temperature observed with ALMA Band 8 is $\sim100~{\rm K}$, while the dust temperature observed with ALMA Band 4 is $\sim180$ K. 
In comparison, the midplane temperature is $\sim200$ K.

\begin{figure*}[ht]
\begin{center}
\includegraphics[scale=0.66]{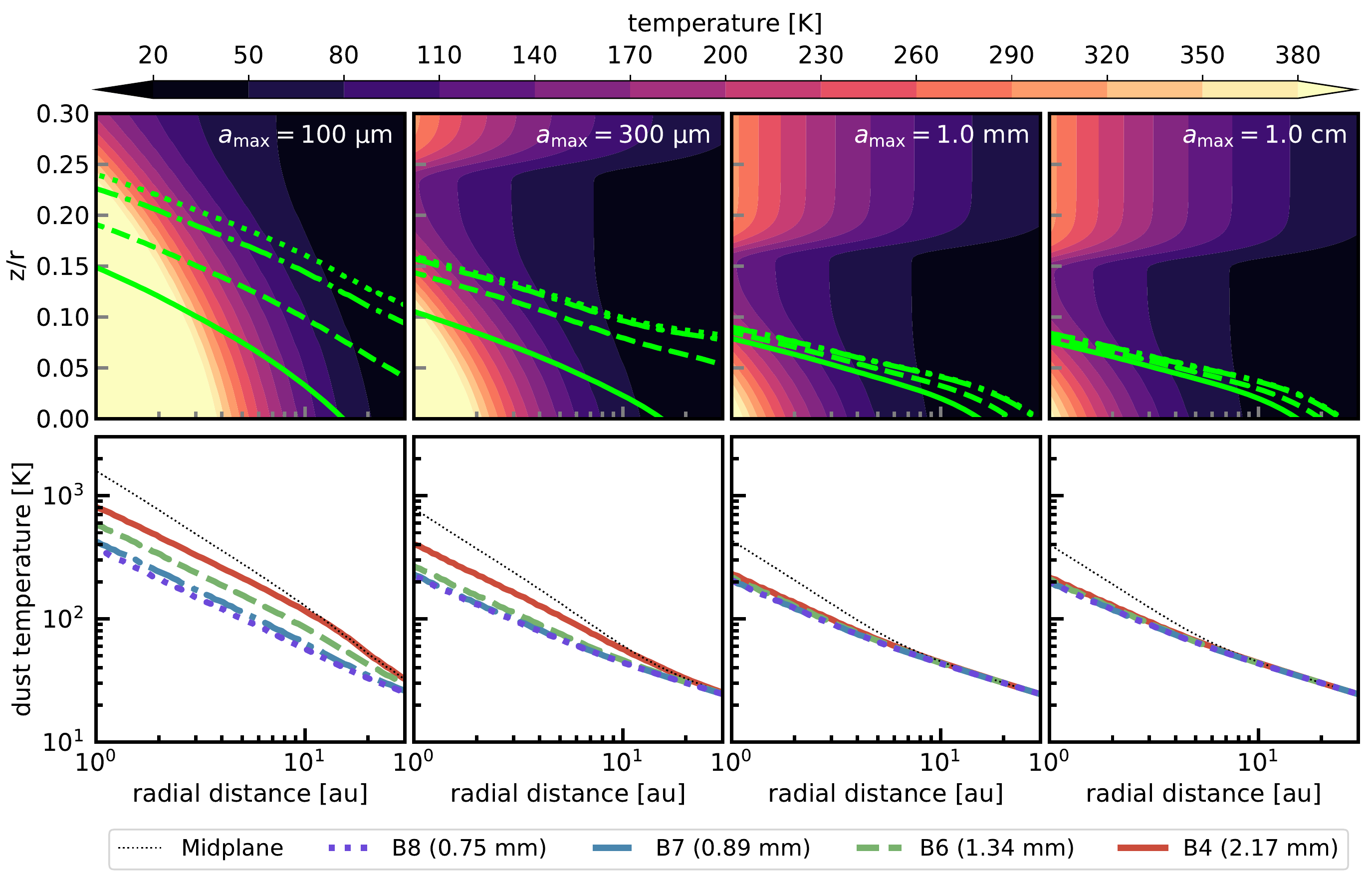}
\caption{
Two-dimensional temperature structure and dust temperature at ALMA surfaces for different values of $a_{\rm max}$.
Top: two-dimensional temperature structure of the fiducial active disk. The light green lines denote the height where the vertical extinction optical depth reaches unity at ALMA Bands 4 (solid), 6 (dashed), 7 (dash-dotted) and 8 (dotted).
bottom: dust temperature at the height where the vertical extinction optical depth at the ALMA wavelengths reaches unity. The radial profile are not beam-convolved.
The dust settling is not taken into account.
The dust is assumed to be compact with a size distribution of $p_{\rm d}=3.5$.
The dust surface density is set so that $\tau_{\rm 10,B4}=3$.
}
\label{fig:2dtemp_amax}
\end{center}
\end{figure*}

Figure \ref{fig:2dtemp_10um} (bottom panels) also compares the brightness temperature profile obtained from the radiative transfer simulation with the observed values.
The bottom-left panel of Figure \ref{fig:2dtemp_10um} corresponds to the top-left panel in Figure \ref{fig:1dtemp_noacc}.
The brightness temperature of the fiducial active disk is more than two times higher than that of the passive disk.
Furthermore, a wavelength-dependent brightness temperature is seen within $\sim10$ au, which is not seen in the observations.
This is because of the vertical temperature gradient shown in the top and middle panels in Figure \ref{fig:2dtemp_10um}.
The brightness temperature is independent of the inclusion of scattering because of the negligible scattering albedo.
Based on these, we conclude that the maximum dust size of 10 ${\rm \mu m}$ is unlikely regardless of the efficiency of accretion heating.

\subsubsection{Fiducial active disks} \label{sec:active}

Figure \ref{fig:2dtemp_amax} shows the two-dimensional temperature structure of the fiducial active disk, considering different values of $a_{\rm max}$.
We see that the temperature of the inner disk region is higher for smaller $a_{\rm max}$ because of the higher vertical optical depth for dust thermal emission ($\tau_{z}$).
For $a_{\rm max}\lesssim300~{\rm \mu m}$, the dust temperature at the ALMA surfaces is higher at longer wavelengths, as shown in Section \ref{sec:10um}.
In contrast, For $a_{\rm max}\gtrsim1~{\rm mm}$, the dust temperature is not sensitive to the observing wavelengths.
This is because the extinction opacity of large dust ($\gtrsim1$ mm) is less sensitive to the ALMA wavelength than that of small dust (see Fig \ref{fig:opac}), and hence all ALMA observations trace similar disk heights.
The difference in the dust temperature is the most significant at $a_{\rm max}\sim100~{\rm \mu m}$ because of its steep slope in the extinction opacity.
Although the dust temperature is nearly independent on the observing wavelengths when $a_{\rm max}\gtrsim1~{\rm mm}$, it remains lower than the midplane temperature at $\lesssim6$ au. 

The temperature structure is almost identical for models with $a_{\rm max}=1~{\rm mm}$ and $1~{\rm cm}$.
This similarity arises because we ensure that the vertical extinction optical depth at ALMA Band 4 ($\tau_{\rm 10,B4}$) is constant across all models.
In the regime of $a_{\rm max}\gtrsim1~{\rm mm}$, both the millimeter extinction opacity and the Rosseland-mean opacity decrease with increasing dust radius in the same manner ($\propto a_{\rm max}^{-1/2}$).
To keep $\tau_{\rm 10,B4}$ to be constant, the dust surface density is inversely proportional to the millimeter extinction opacity.
Consequently, despite the decrease in the Rosseland-mean opacity as the dust radius increases, the infrared optical depth $\tau_{\rm z}$, which governs the magnitude of accretion heating, remains unchanged when $\tau_{\rm 10,B4}$ is held constant.

\begin{figure*}[t]
\begin{center}
\includegraphics[scale=0.45]{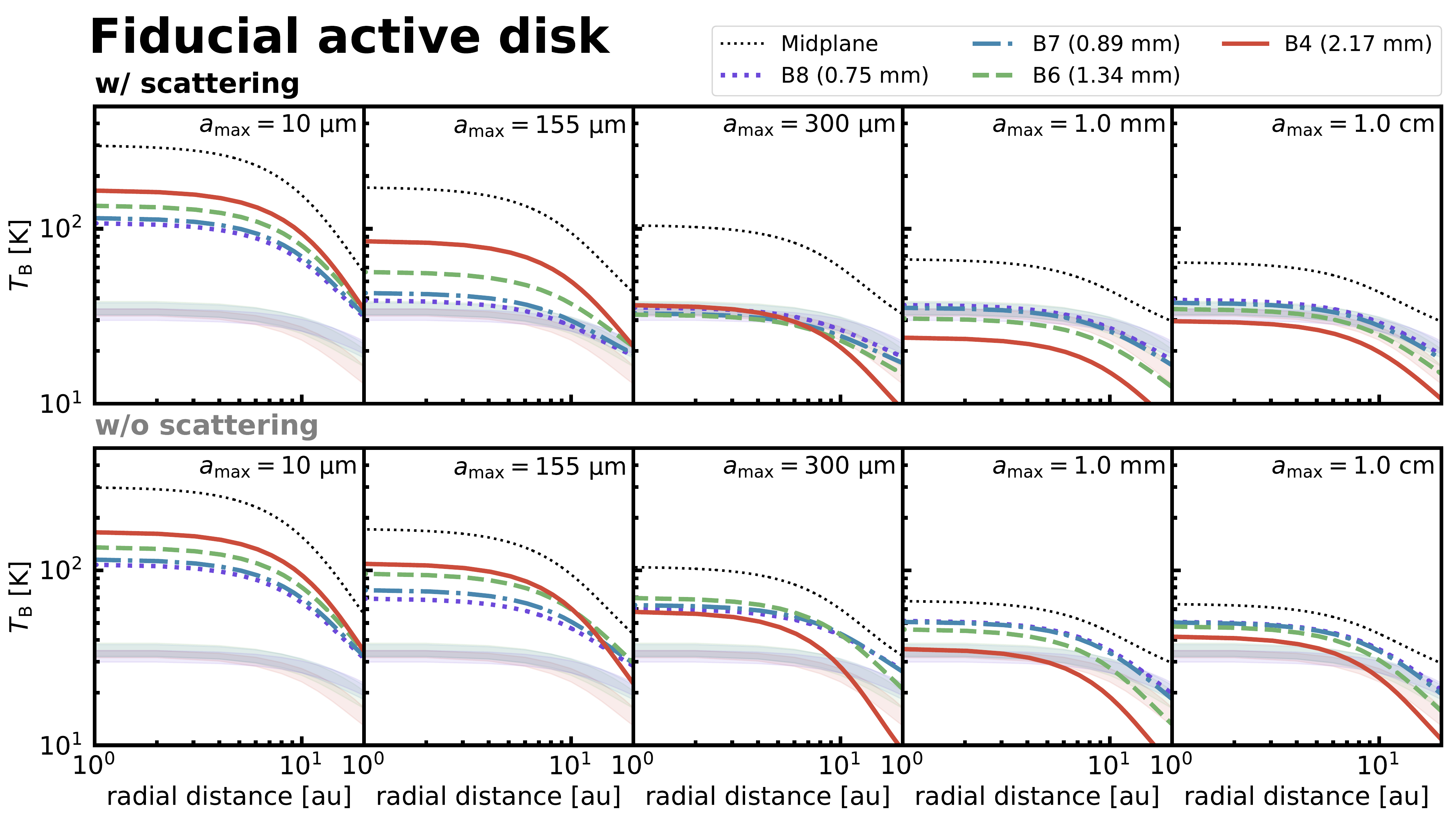}
\caption{
Same as Figure \ref{fig:1dtemp_noacc} but for the fiducial active disk.
}
\label{fig:1dtemp_fid}
\end{center}
\end{figure*}

Figure \ref{fig:1dtemp_fid} shows the brightness temperature profile of the fiducial active disk model.
In contrast to the case of $a_{\rm max}=10~{\rm \mu m}$, when $a_{\rm max}\gtrsim100~{\rm \mu m}$, scattering is no longer negligible and reduces the brightness temperature which makes the interpretation of the brightness temperature more complicated.

For $a_{\rm max}=155~{\rm \mu m}$, the model brightness temperatures are significantly higher than the observed values.
The millimeter-wave scattering makes the brightness temperatures lower than that of no-scattering limit.
The model with $a_{\rm max}=300~{\rm \mu m}$ shows similar brightness temperatures with the observations if scattering is included.
The brightness temperatures with scattering are $\sim2$ times lower than that without scattering.
Furthermore, the midplane temperature is $\sim3$ time higher than the brightness temperatures.
This clearly demonstrates that, even if the disk is fully optically thick, the observed brightness temperature can be much lower than the true midplane temperature because of the combined effect of scattering and vertical temperature gradient (see also \citealt{SL20}).

\begin{figure}[ht]
\begin{center}
\includegraphics[scale=0.53]{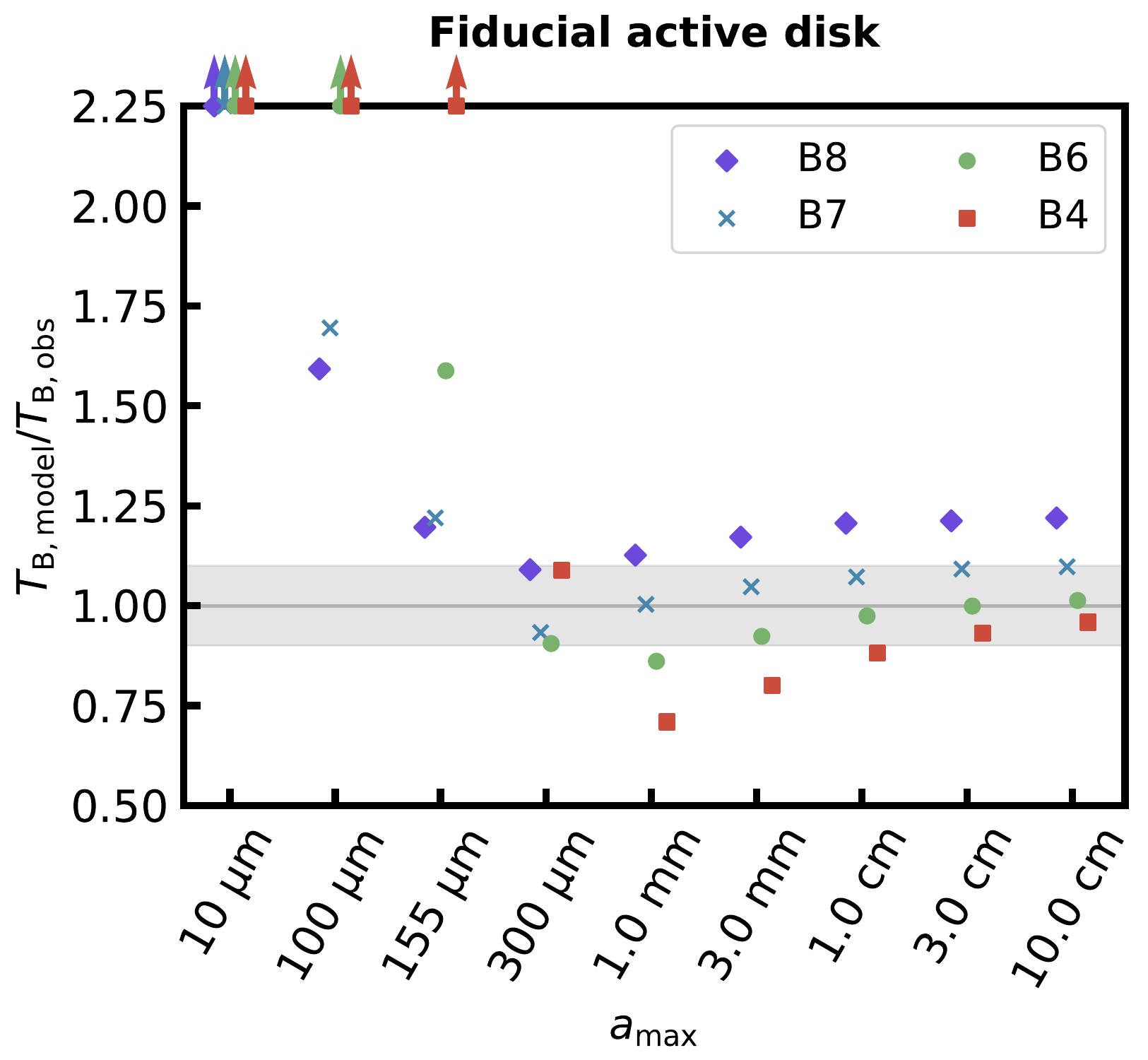}
\caption{
Same as Figure \ref{fig:passive_center} but for the fiducial active disk.
}
\label{fig:active_center}
\end{center}
\end{figure}

When $a_{\rm max}$ is of the order of 1 mm, the brightness temperature at Band 4 is significantly lower than the observed value because of scattering even if the accretion heating takes place (see also Figure \ref{fig:active_center}).
In contrast, when $a_{\rm max}$ is greater than $\sim$ 1 cm, the model brightness temperatures are similar to the observations as seen for $a_{\rm max}=300~{\rm \mu m}$.
This is because the effective albedo at ALMA Band 4 decreases as $a_{\rm max}$ increases in the regime of $a_{\rm max}\gtrsim1~{\rm mm}$.
In this case, the midplane temperature is $\sim2$ times higher than the observed brightness temperatures.

The central brightness temperature of the fiducial active disks is shown in Figure \ref{fig:active_center}.
In the small dust regime ($<300~{\rm \mu m}$), the model brightness temperature are too high to account for the observations because of the efficient accretion heating.
While the passive disk model with $a_{\rm max}\sim300~{\rm \mu m}$ underestimates the brightness temperature at ALMA Bands 4-7, the fiducial active disk model with $a_{\rm max}\sim300~{\rm \mu m}$ aligns well with the observations.
This is attributed to the enhancement of brightness temperature by accretion heating, particularly at longer wavelengths.
Large grain models ($\gtrsim$ few cm) are roughly consistent with the observations, although they tend to overestimate the brightness temperature at Band 8.

\subsection{Effect of dust settling, size distribution and porosity}
\begin{figure*}[t]
\begin{center}
\includegraphics[scale=0.66]{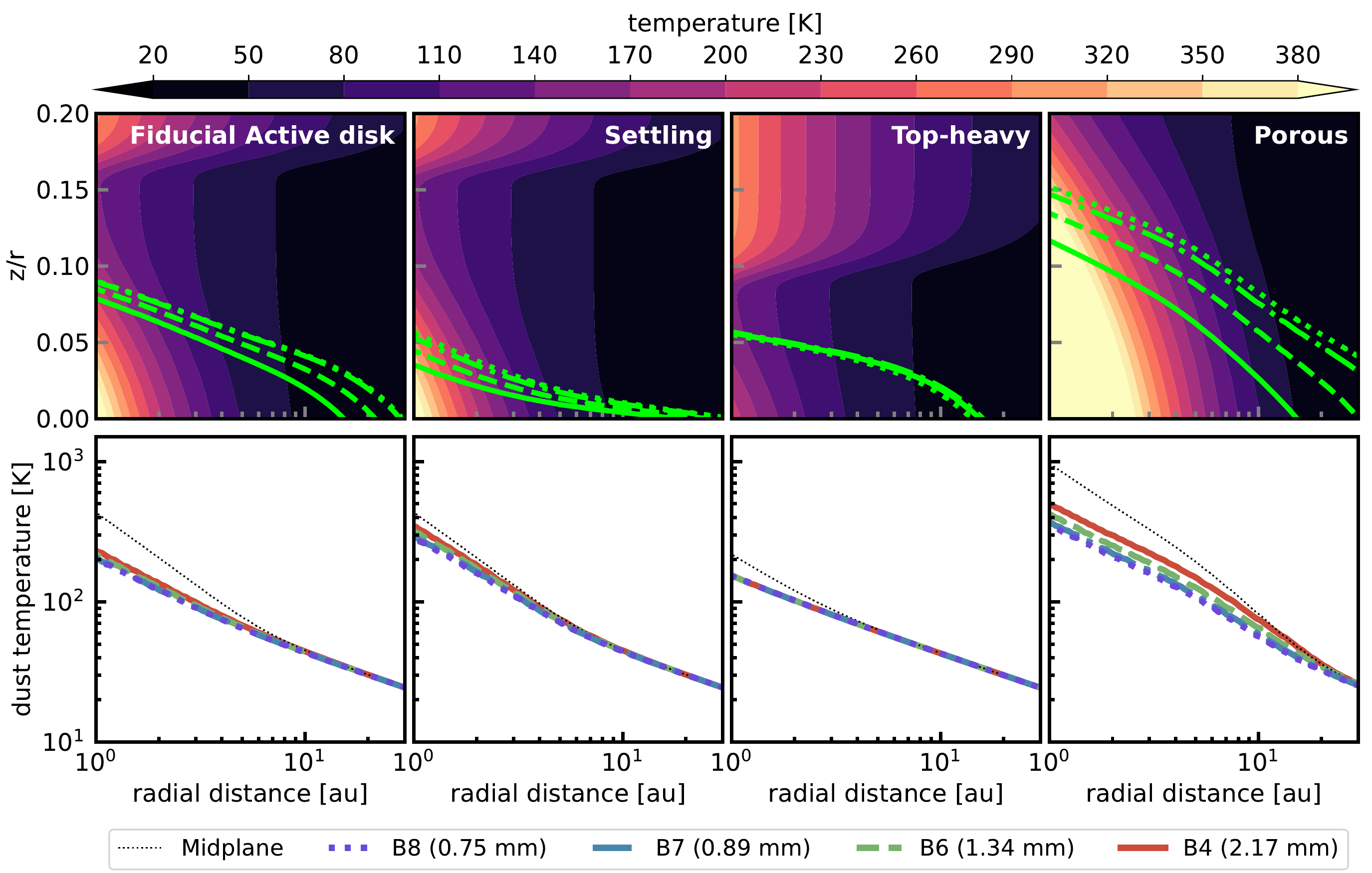}
\caption{
Two-dimensional temperature structure and dust temperature at ALMA surfaces for different disk models.
Top: two-dimensional temperature structure of the fiducial active disk model (left), Settling model ($\alpha_{\rm t}=10^{-4}$, second left), Top-heavy model (second right) and Porous model (right) . The light green lines denote the height where the vertical extinction optical depth reaches unity at ALMA Bands 4 (solid), 6 (dashed), 7 (dash-dotted) and 8 (dotted).
The maximum dust size $a_{\rm max}$ is 1 mm in the all models.
Bottom: dust temperature at the height where the vertical extinction optical depth at the ALMA wavelengths reaches unity. 
The radial profile is not convolved.
}
\label{fig:2dtemp_models}
\end{center}
\end{figure*}

In this section, we explore the influence of dust settling, dust size distribution, and dust porosity on the brightness temperatures of the active disk. 
Figure \ref{fig:2dtemp_models} shows the two-dimensional temperature structure of the fiducial active disk (same as the second right column of Figure \ref{fig:2dtemp_amax}), settling model, top-heavy model and porous dust model.
The maximum dust size is set to 1 mm.
We discuss the effect of each component in the following subsections.

\subsubsection{Effect of dust settling} \label{sec:set}
As dust grains grow larger, they are expected to settle down towards the midplane of the disk. 
If dust settling takes place, the effective dust size that contributes to the millimeter emission decreases because larger grains can be hidden within the optically thick midplane layer \citep{Ueda+21}.

Figure \ref{fig:2dtemp_models} compares the dust temperature at the ALMA surfaces of the models without (left) and with (second left) dust settling with $\alpha_{\rm t}=10^{-4}$.
We find that dust settling effectively increases the temperature of dust observed with ALMA.
This behavior can be explained as follows:
The efficiency of accretion heating is primarily determined by the distribution of small grains, whereas the height of the ALMA surfaces is predominantly influenced by the distribution of large grains. Due to settling, larger grains tend to selectively settle towards the midplane, causing the ALMA surfaces to descend more rapidly towards the midplane compared to the isotherm height of the active disk (as shown in the top rows of Figure \ref{fig:2dtemp_models}). This allows ALMA to observe a hotter layer.
\begin{figure}[ht]
\begin{center}
\includegraphics[scale=0.53]{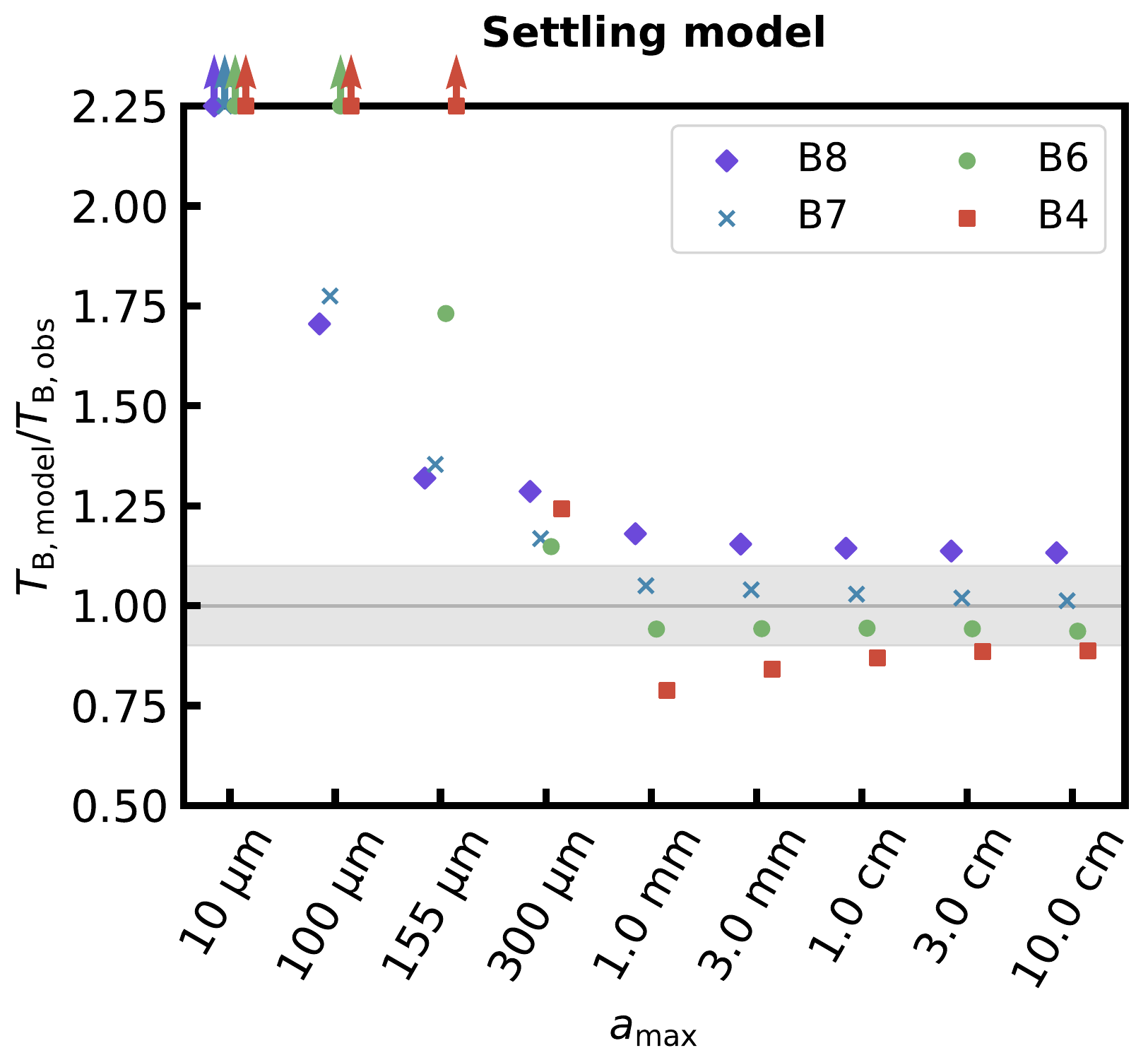}
\caption{
Same as Figure \ref{fig:passive_center} but for the settling model with $\alpha_{\rm t}=10^{-4}$.
}
\label{fig:settling_center}
\end{center}
\end{figure}

Figure \ref{fig:settling_center} shows the central brightness temperature of the beam-convolved images of the settling model. 
The radial brightness temperature profile of each model is shown in the appendix \ref{sec:app2} (see Figure \ref{fig:1dtemp_set}).
For $a_{\rm max}\lesssim3~{\rm mm}$, the brightness temperature is systematically higher than that of the fiducial active disk model (i.e., without settling).
This is because of the effect explained above.
Particularly for $a_{\rm max}=300~{\rm \mu m}$, the settling model overpredicts the brightness temperatures because of the higher temperature of the dust responsible for the emission.
We note that the settling model with $a_{\rm max}=300~{\rm \mu m}$ reasonably explains the observations if $\alpha_{\rm t}=10^{-3}$ because dust settling is less effective compared to $\alpha_{\rm t}=10^{-4}$.
We also note that if the disk is active, the turbulence strength for dust mixing may not be significantly lower than that for accretion ($\alpha_{\rm acc}$; Equation \eqref{eq:alpha_acc}), implying that $\alpha_{\rm t}=10^{-3}$ may be more likely for the active disk model (see also Section \ref{sec:sigmad}).

For $a_{\rm max}\gtrsim3~{\rm mm}$, the brightness temperatures of the settling model are not simply higher than those of the model without settling.
The temperature of dust observed with ALMA is higher for the models with settling compared to those without settling.
However, the differential settling makes effective dust size smaller, which increases the scattering albedo and hence reduces the emergent intensity at the ALMA wavelengths.
The emergent intensity, and hence the brightness temperature, is determined by the balance between these two effects.

Overall, large grain models ($a_{\rm max}\gtrsim3~{\rm mm}$) predict the brightness temperatures similar to the observations. 
Compared to the no-settling models (Figure \ref{fig:active_center}), the settling model is more consistent with the observations if $a_{\rm max}\gtrsim3~{\rm mm}$. 
We note that if $\alpha_{\rm t}=10^{-3}$, dust settling has no significant impact on the brightness temperatures when $a_{\rm max}\lesssim1 {\rm mm}$ (see also Section \ref{sec:sigmad}).

\subsubsection{Effect of dust size distribution}
The dust size distribution plays a critical role in both the millimeter thermal emission and the efficiency of accretion heating.
In our fiducial case, we assume a slope of $p_{\rm d}=3.5$ for the dust size distribution. 
However, it is important to note that the dust size distribution in protoplanetary disks is uncertain and is influenced by dust coagulation and fragmentation processes.
In particular, if dust fragmentation is not efficient, the dust size distribution can exhibit a more top-heavy nature, meaning it is dominated by larger grains, rather than following a slope of $p_{\rm d}=3.5$ (e.g., \citealt{BOD11}).

\begin{figure}[ht]
\begin{center}
\includegraphics[scale=0.53]{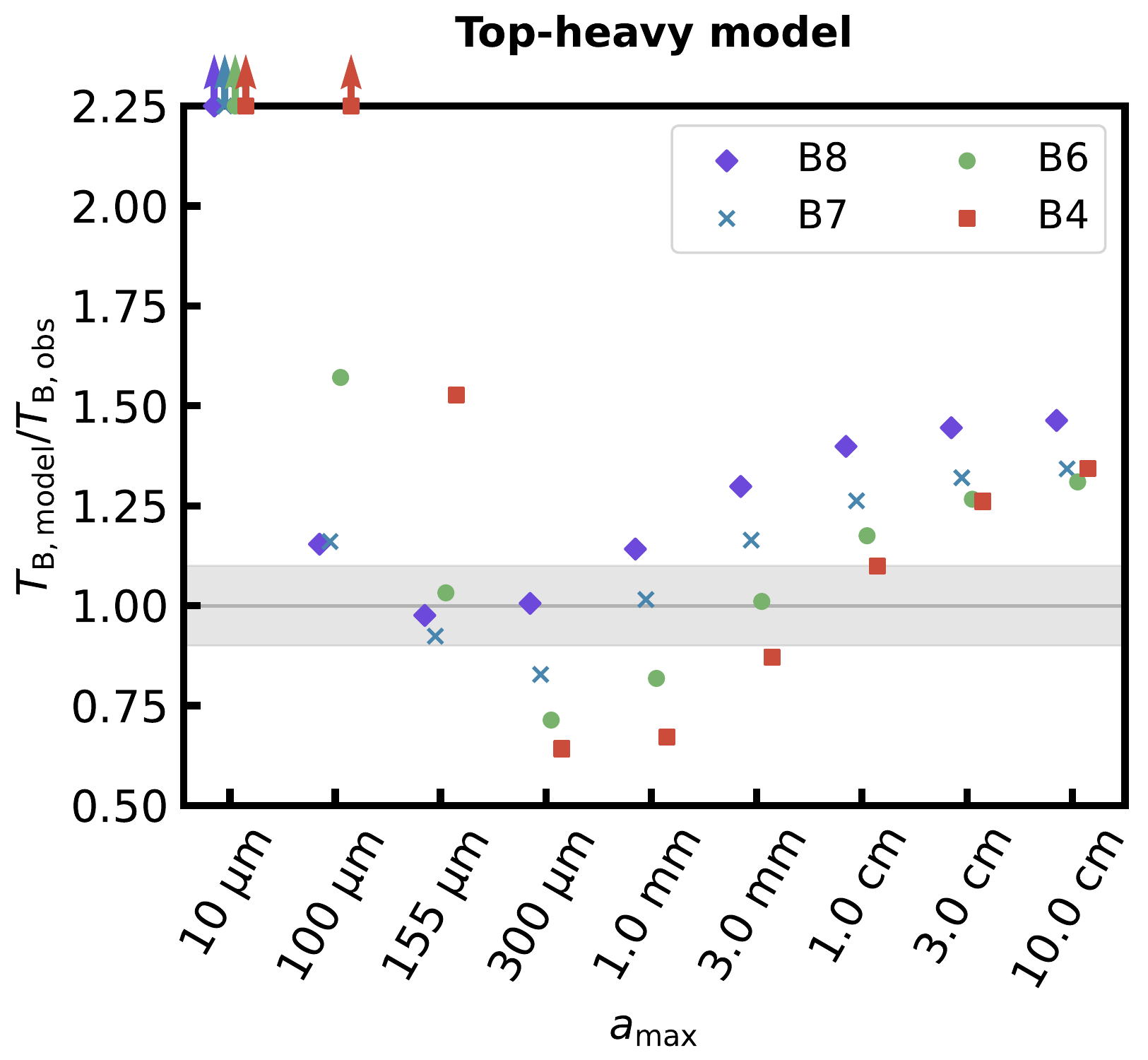}
\caption{
Same as Figure \ref{fig:passive_center} but for the top-heavy dust model.
}
\label{fig:topheavy_center}
\end{center}
\end{figure}

In the top-heavy model, we adopt a dust size distribution with a slope of $p_{\rm d}=2.5$ instead of 3.5.
As shown in Figure \ref{fig:2dtemp_models}, the top-heavy dust size distribution leads to a cooler disk region near the midplane compared to the fiducial active disk.
This is because the dust mass is more dominated by large grains which have lower surface-to-mass ratio and hence accretion heating is less efficient.

Figure \ref{fig:topheavy_center} shows the central brightness temperature of the beam-convolved images of the top-heavy model. 
The radial brightness temperature profile of each model is shown in the appendix \ref{sec:app2} (see Figure \ref{fig:1dtemp_pd2.5}).
Although the dust temperature is lower than that of fiducial active disk model, the brightness temperature does not simply follow the dust temperature.
In the regime of $a_{\rm max}\lesssim1~{\rm mm}$, scattering is more efficient than the case of $p_{\rm d}=3.5$ because mm-sized grains have higher scattering albedo than micron-sized grains at ALMA wavelengths.
Therefore, for $a_{\rm max}\lesssim1~{\rm mm}$, the brightness temperatures of the top-heavy model are lower compared to the fiducial active disk model.
In contrast, in the regime of $a_{\rm max}\gtrsim1~{\rm mm}$, scattering is less efficient than the case of $p_{\rm d}=3.5$ because cm-sized or larger grains have lower scattering albedo than mm-sized grains at ALMA wavelengths.
Therefore, for $a_{\rm max}\gtrsim1~{\rm mm}$, the brightness temperatures of the top-heavy model are higher compared to the fiducial active disk model.
In the middle of these two regimes ($a_{\rm max}\sim1~{\rm mm}$), the brightness temperature is not sensitive to $p_{\rm d}$.
Overall, the top-heavy models do not appear to be consistent with the observations.

\subsubsection{Effect of porosity} \label{sec:por}
In the previous sections, we focused on studying the brightness temperature of disks with compact dust grains.
However, it is important to consider that dust in protoplanetary disks can be porous due to the pairwise collisional growth (e.g., \citealt{Okuzumi+12}).

\begin{figure}[ht]
\begin{center}
\includegraphics[scale=0.47]{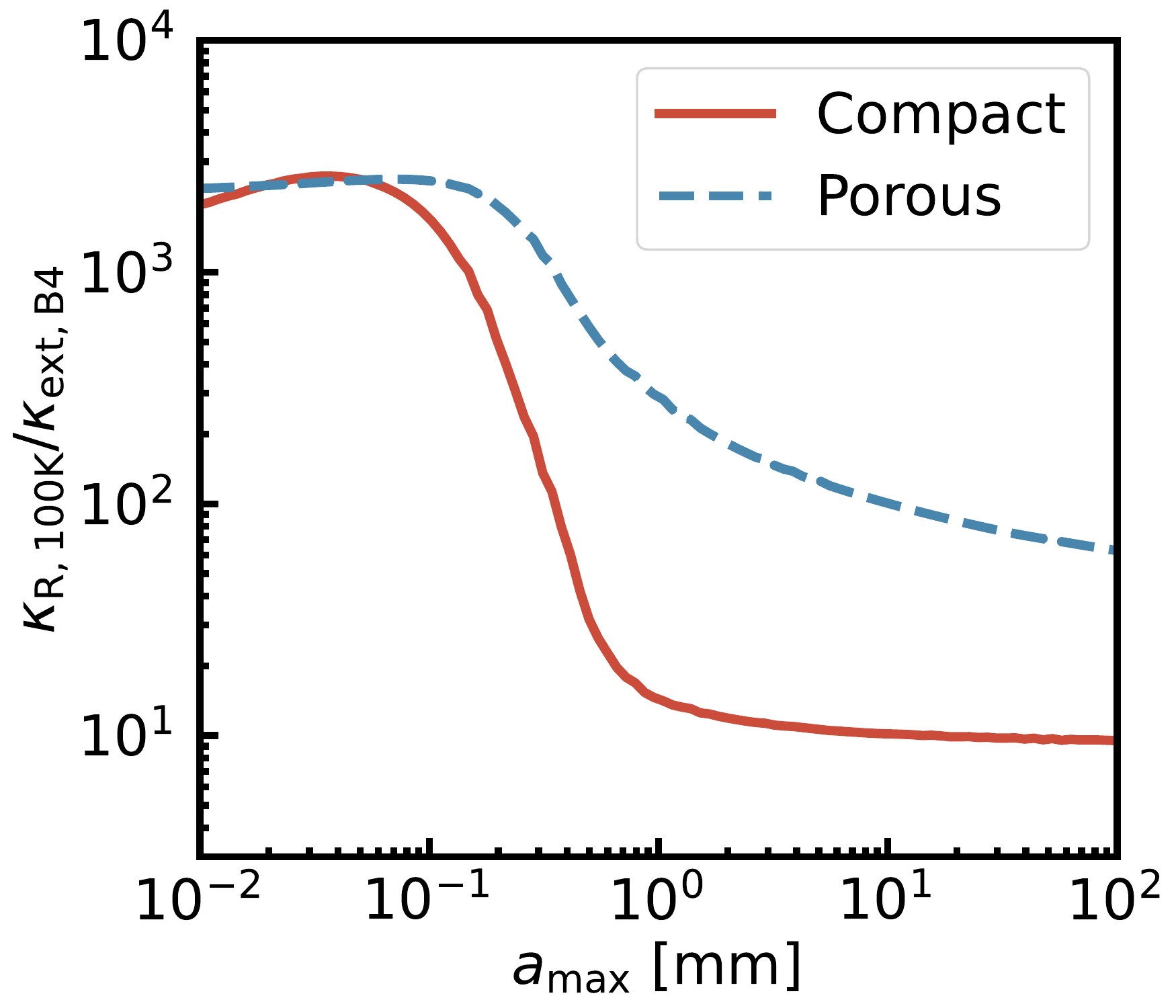}
\caption{
Ratio between the Rosseland-mean opacity (at 100K) and the extinction opacity at ALMA Band 4.
The red solid and blue dashed lines denote the compact and porous dust, respectively.
}
\label{fig:opac-ratio}
\end{center}
\end{figure}

In Figure \ref{fig:2dtemp_models}, we compare the dust temperature between the fiducial active disk model (left) and the porous dust model (right). 
We see that the accretion heating is more efficient in the porous dust model. 
This is because, for a given optical depth at ALMA wavelengths (e.g., $\tau_{\rm B4}$), the optical depth for the dust thermal emission ($\tau_{\rm z}$) is higher for porous dust.
Figure \ref{fig:opac-ratio} shows the ratio between the Rosseland-mean opacity (at 100K) and the extinction opacity at ALMA Band 4.
For $a_{\rm max}\gtrsim100~{\rm \mu m}$, porous dust has higher $\kappa_{\rm R,100K}/\kappa_{\rm ext,B4}$. 
This means that, for a given optical depth at ALMA wavelength, porous dust has larger vertical optical depth for accretion heating and hence the accretion heating is more efficient compared to compact dust.

\begin{figure}[ht]
\begin{center}
\includegraphics[scale=0.53]{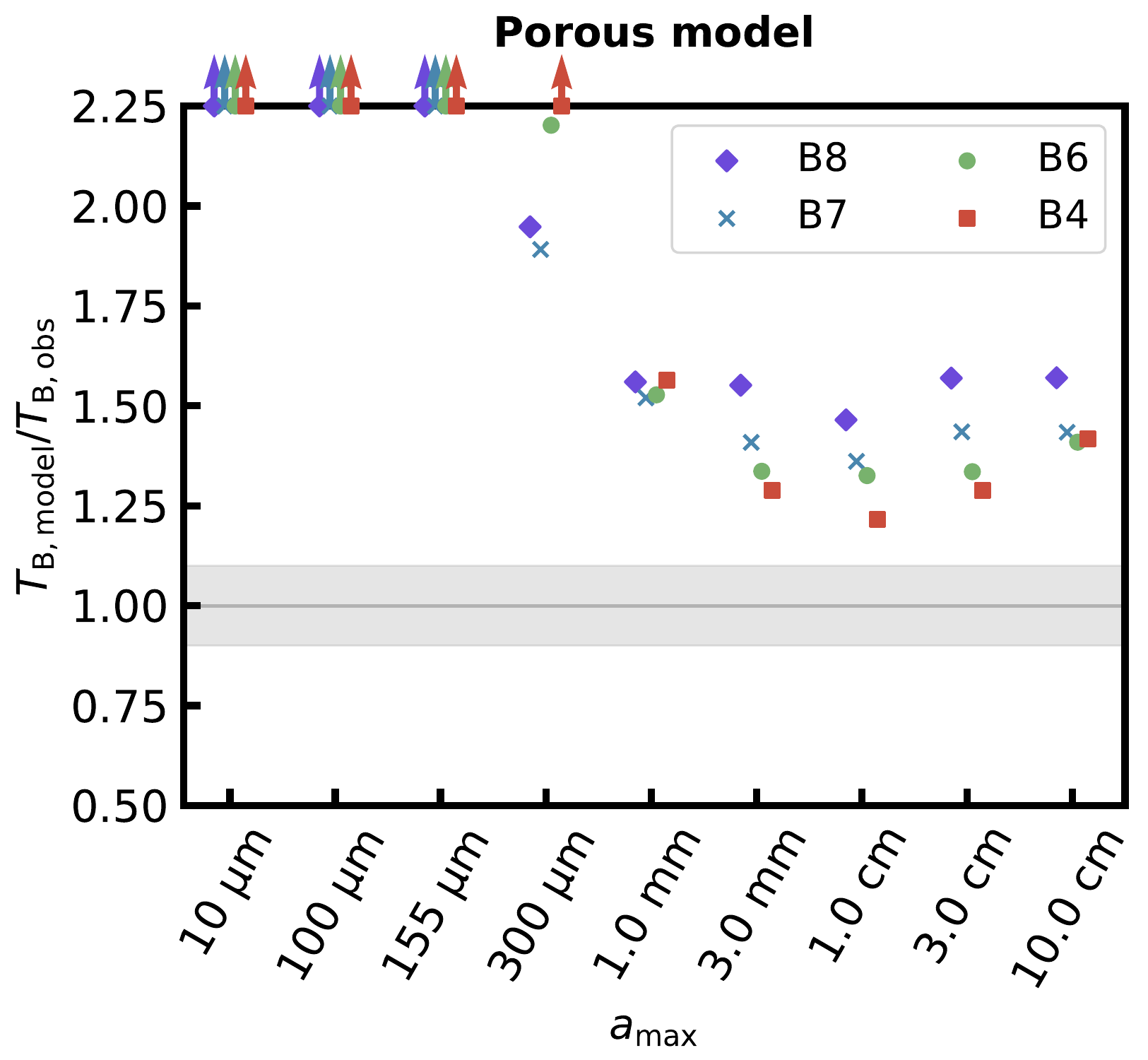}
\caption{
Same as Figure \ref{fig:passive_center} but for the porous dust model.
}
\label{fig:porous_center}
\end{center}
\end{figure}

Figure \ref{fig:porous_center} shows the central brightness temperature of the convolved images of the porous dust model. 
The radial brightness temperature profile of each model is shown in the appendix \ref{sec:app2} (see Figure \ref{fig:1dtemp_porous}).
If dust is porous, the brightness temperature is higher than that of compact dust model regardless of the dust size.
We see that the porous dust models predict $\gtrsim1.5$ times higher brightness temperatures than the observations.
We also confirm that the brightness temperatures of porous dust model are still higher even if $p_{\rm d}=2.5$.

\section{Discussion} \label{sec:discussion}

\subsection{Dust size in the inner region of the CW Tau disk}
Our study reveals that the maximum dust size in the inner region ($\lesssim10$ au) of the CW Tau disk is preferred to be $\sim$150--300 ${\rm \mu m}$ (small grain solution) or larger than a few centimeters (large grain solution).
Distinguishing between these two scenarios is essential for gaining insights into grain growth and temperature structure of the disk.

The CW Tau disk shows the strong scattering-induced polarization at ALMA Band 7 with a polarization degree of $\gtrsim1$\% \citep{Bacciotti+18}.
This indicates that the thermal emission at ALMA Band 7 is primarily contributed by dust grains with a radius of $\lambda/2\pi\sim140~{\rm \mu m}$ (\citealt{Kataoka+15}), although the angular resolution of the polarimetric observations ($\sim$0\farcs2) is worse than ours (0\farcs1).
This implies that the small grain solution is more consistent with the polarimetric observations than the large grain solution.

The effective observed-dust size can be smaller than the true maximum dust size if differential dust settling takes place (\citealt{SL20,Ueda+21}), which potentially makes the large grain solution matches with the polarimetric observation.
However, it is important to note that achieving the observed high polarization degree ($\gtrsim1$\%) with $a_{\rm max}$ larger than $\sim1$ mm is unlikely, even under the assumption of extremely low turbulence strength \citep{Ueda+21}.

The polarization degree also depends on the detailed structure of dust grains.
Recent laboratory experiment suggests that non-spherical irregular shape dust can produce strong scattering-induced polarization even if the dust size is significantly larger than the observing wavelength \citep{Lin+23}.
If this is the case, the polarization degree of non-spherical grains is less sensitive to the observing wavelength than the compact spherical grains.
Furthermore, inclusion of small amount of porosity ($p\lesssim0.9$) also makes the polarization degree less sensitive to the dust size (\citealt{Tazaki+19,Zhang+23}).
However, we note that, if the porosity is as high as 0.9, accretion heating is so efficient that the model brightness temperatures exceed the observed values.
Therefore, if $p\gtrsim0.9$, some sort of reduction of accretion heating due to, e.g., wind-driven accretion \citep{Mori+19,Kondo+22}, is required.
Future multi-wavelength polarimetric observations would help us constrain the detailed dust structure.

\subsection{Surface densities and effective accretion $\alpha$} \label{sec:sigmad}
In this section, we discuss the expected dust and gas surface densities, as well as the effective turbulence strength for the accretion of the CW Tau disk.

\begin{figure}[ht]
\begin{center}
\includegraphics[scale=0.5]{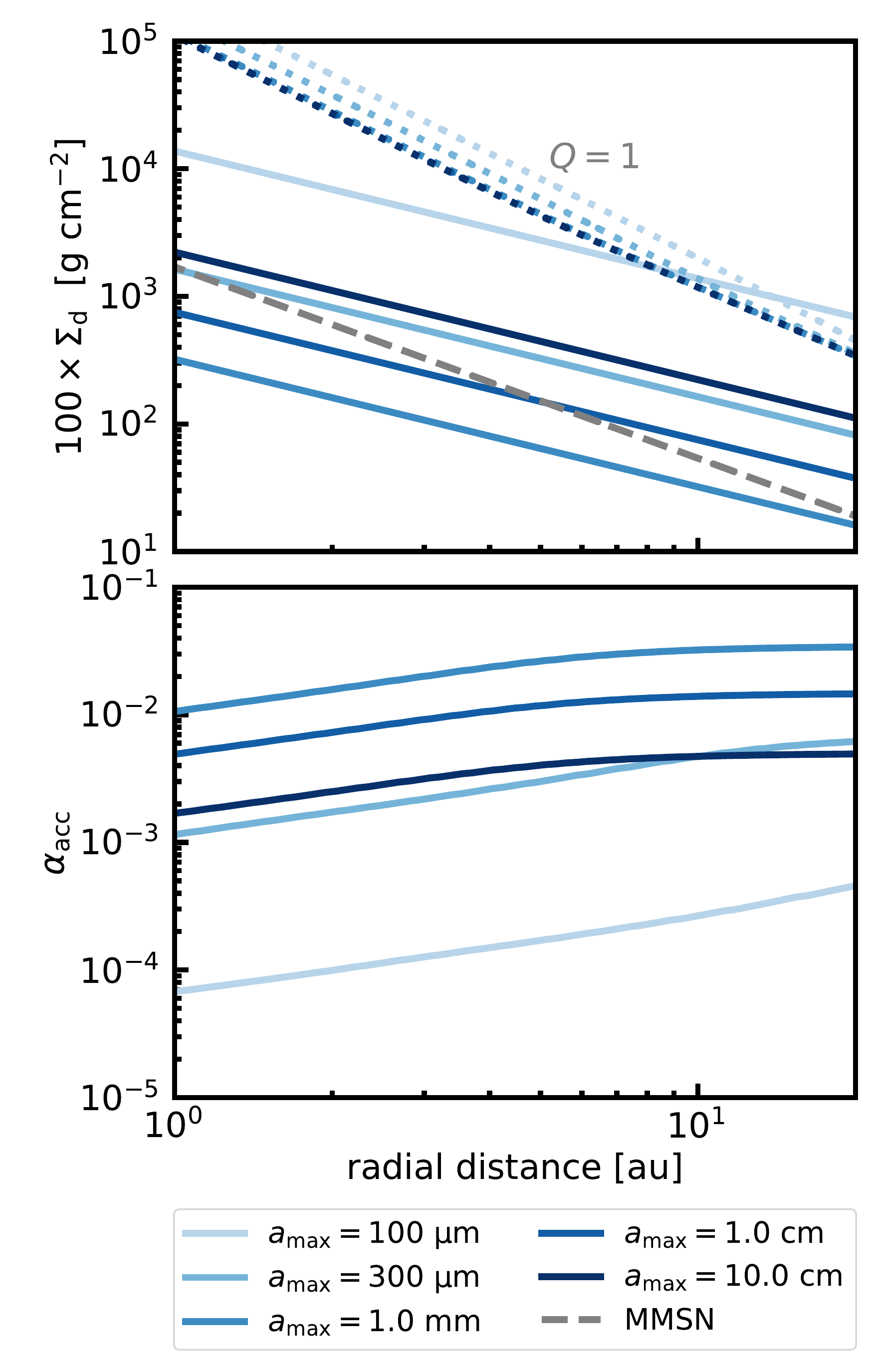}
\caption{
Gas surface density (hundred times of the dust; top) and effective turbulence strength for the disk gas accretion (bottom) estimated from  our fiducial active disk models. 
The gray dashed line denotes the gas surface density of the minimum mass solar nebula (MMSN; \citealt{Hayashi81}).
The dotted lines denote the criteria for the disk to be gravitationally unstable ($Q=1$).
}
\label{fig:Sigmad}
\end{center}
\end{figure}

Figure \ref{fig:Sigmad} shows the gas surface density of our fiducial active disk models(assuming hundred times of the dust).
The gas surface density of the minimum mass solar nebula (MMSN), $1700(r/{\rm au})^{-1.5}~{\rm g~cm^{-2}}$ \citep{Hayashi81}, is also shown for comparison.
We also plot the criterion for the disk to be graivationally unstable (\citealt{Toomre1964}):
\begin{eqnarray}
Q\equiv\frac{\Omega_{\rm K}c_{\rm s}}{\pi G \Sigma_{\rm g}}=1.
\end{eqnarray}

In general, our model disks are expected to be gravitationally stable within the region of focus, approximately $\lesssim10$ au. 
If $a_{\rm max}=100~{\rm \mu m}$, the gas surface density reaches $Q=1$ around $\sim10$ au.
If $a_{\rm max}$ were $10~{\rm \mu m}$, the disk mass would be $\sim1.5$ times higher; however, as already shown in Section \ref{sec:comparison}, the model with $a_{\rm max}=10~{\rm \mu m}$ fails to reproduce the observed brightness temperatures.
On the other hand, for $a_{\rm max}\gtrsim300~{\rm \mu m}$, the gas surface density is estimated to be much lower than the $Q=1$ line within 20 au.
Moreover, the estimated gas surface density aligns closely with the MMSN model within 10 au for $a_{\rm max}\gtrsim300~{\rm \mu m}$.
This indicates that the CW Tau disk has enough capability to form a planetary system similar to our solar system.

Figure \ref{fig:Sigmad} also shows the effective turbulence strength for the disk accretion;
\begin{eqnarray}
\alpha_{\rm acc}=\frac{\dot{M}}{3\pi\Sigma_{\rm g}c_{\rm s}h_{\rm g}}. \label{eq:alpha_acc}
\end{eqnarray}
For the given accretion rate of $4\times10^{-8}M_{\odot}~{\rm yr^{-1}}$, the effective turbulence strength accounting for the disk accretion needs to be $\sim10^{-4}$ for $a_{\rm max}=100~{\rm \mu m}$, while it is $\sim 10^{-3}$--$10^{-2}$ for $a_{\rm max}\gtrsim300~{\rm \mu m}$.
The estimated $\alpha_{\rm acc}$ falls into a reasonable range of $\alpha_{\rm acc}=10^{-4}$--$10^{-2}$, although the high $\alpha_{\rm acc}$ ($\sim10^{-2}$) found for the models of $a_{\rm max}=1$--10 mm is higher than the typical value estimated from the observed disk lifetime in various star-forming regions (\citealt{Manara+22}, but see also \citealt{Hartmann+98}).

The model with $a_{\rm max}=300~{\rm \mu m}$ yields $\alpha_{\rm acc}\sim10^{-3}$.
As shown in Section \ref{sec:active} and \ref{sec:set}, the model with $a_{\rm max}=300~{\rm \mu m}$ can account for the observations if vertical dust settling is ignored but cannot if dust settling takes place with $\alpha_{\rm t}=10^{-4}$.
To further investigate this, we conducted a simulation of the settling model with $\alpha_{\rm t}=10^{-3}$ and confirm that the model with $a_{\rm max}=300~{\rm \mu m}$ reasonably matches the observations if $\alpha_{\rm t}=10^{-3}$.
This suggests that the effective turbulence strength for vertical dust mixing, $\alpha_{\rm t}$, is not much lower than the effective turbulence strength for disk accretion, $\alpha_{\rm acc}$, if $a_{\rm max}=300~{\rm \mu m}$.

\subsection{Predictions for future observations}
Our results show that the observed brightness temperatures at ALMA Bands 4-8 can be explained by either the passive or active disk models.
It is crucial to distinguish these models for understanding the heating mechanisms in the inner region of disks.

\begin{figure}[ht]
\begin{center}
\includegraphics[scale=0.57]{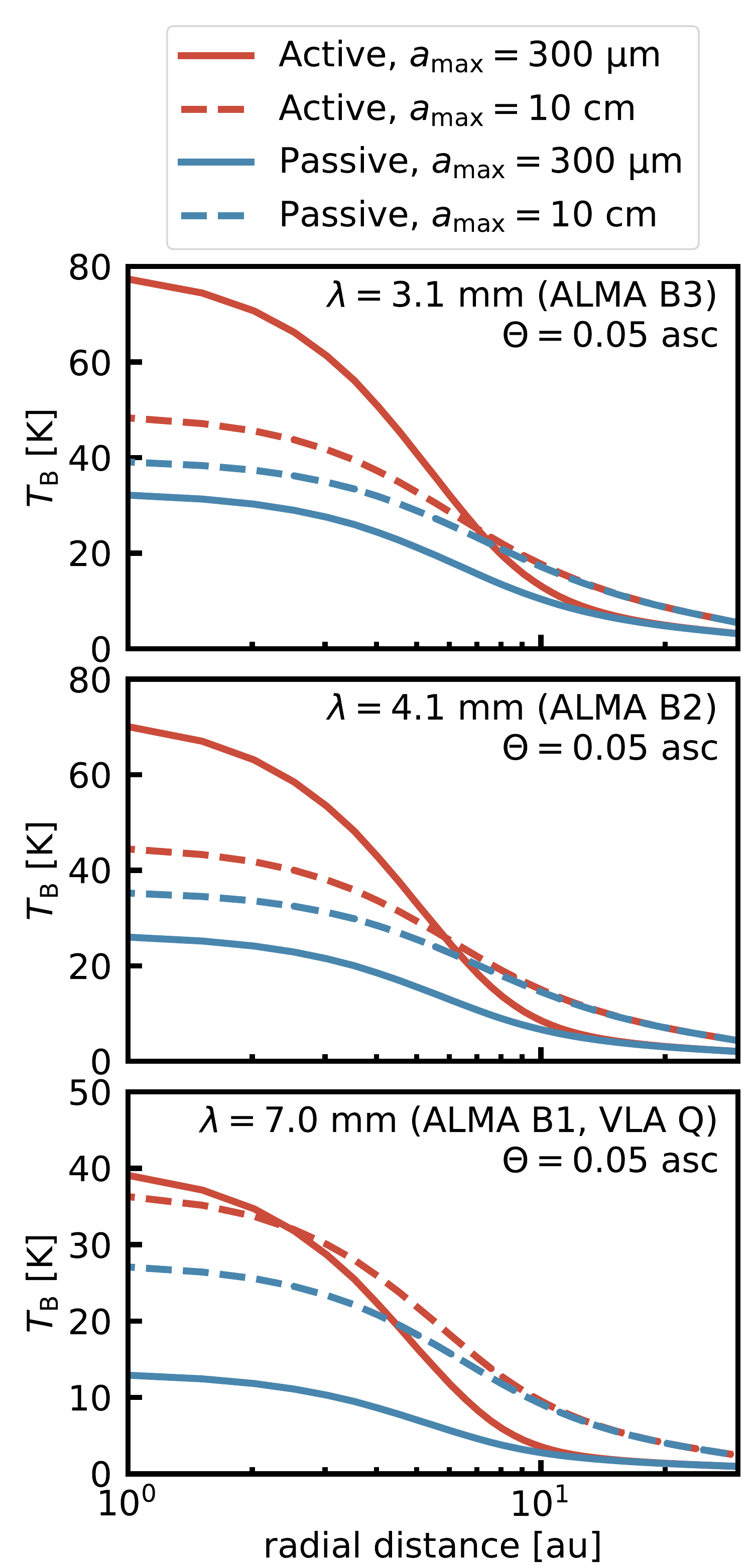}
\caption{
Brightness temperature profile at $\lambda=3.1$, 4.1 and 7.0 mm predicted by our active disk models (solid lines) and passive disk models (dotted lines).
The angular resolution is assumed to be 0\farcs05.
The scattering-induced intensity reduction is included.
}
\label{fig:prediction}
\end{center}
\end{figure}

Figure \ref{fig:prediction} shows the brightness temperatures at $\lambda=3.1$ (ALMA Band 3), 4.1 (ALMA Band 2) and 7.0 mm (ALMA Band 1 or VLA Q band) computed from our models.
The angular resolution is assumed to be 0\farcs05.
The active disk with $a_{\rm max}=300~{\rm \mu m}$ has $\sim$1.5--2 times higher  brightness temperatures at $\lambda=3.1$ and 4.1 mm than the other models.
The peak brightness temperature reaches $\sim80$ K at $\lambda=3.1$ and 4.1 mm.
In contrast, at $\lambda=7.0$ mm, the brightness temperature of the active disk model with $a_{\rm max}=300~{\rm \mu m}$ is not significantly different from that with $a_{\rm max}=10~{\rm cm}$.
This is because the small grain solution has higher temperature but lower optical depth, resulting in the cancellation of these effects.

The passive disk with $a_{\rm max}=300~{\rm \mu m}$ can be distinguished from the other models by using $\lambda=7.0$ mm because it has significantly lower brightness temperature.
This is because the passive disk has lower temperature than the active disk and the optical depth is lower than the large grain model.

If $a_{\rm max}=10~{\rm cm}$, the accretion heating is not so significant that both the passive and active disk predict similar brightness temperature.
However, the difference in those models is potentially distinguishable by ALMA because typical uncertainty in the ALMA flux is 5-10\% which is smaller than the difference in the brightness temperatures predicted by the models.

\begin{figure}[ht]
\begin{center}
\includegraphics[scale=0.57]{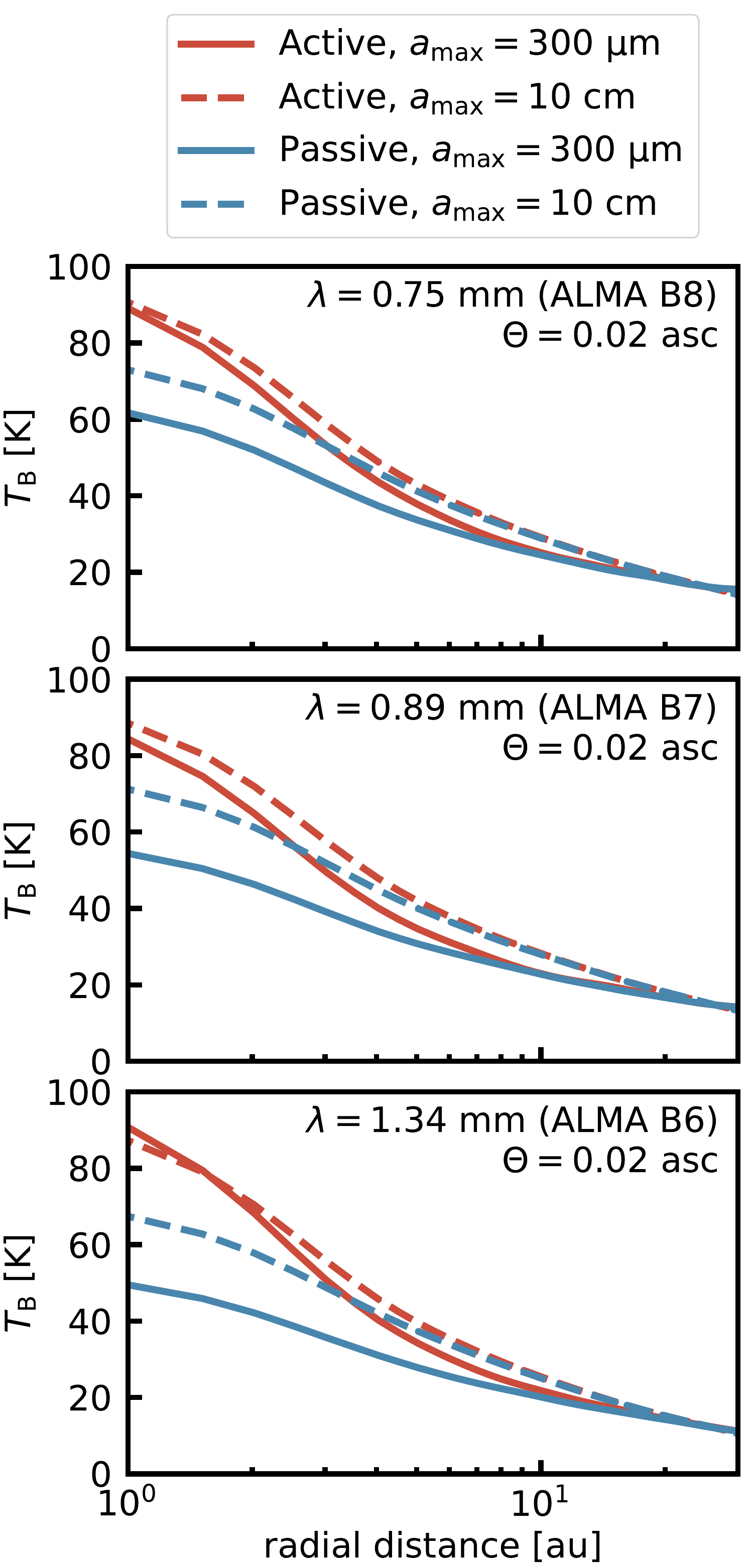}
\caption{
Brightness temperature profile at $\lambda=0.75$, 0.89 and 1.34 mm predicted by our active disk models (solid lines) and passive disk models (dotted lines).
The angular resolution is assumed to be 0\farcs02.
The scattering-induced intensity reduction is included.
}
\label{fig:prediction_short}
\end{center}
\end{figure}

Even though the finest angular resolution of current facilities (ALMA and VLA) at $\lambda\gtrsim3~{\rm mm}$ is limited to $\sim$ 0\farcs05, it can achieve $\sim$ 0\farcs02 at shorter wavelengths.
Figure \ref{fig:prediction_short} shows the brightness temperature at $\lambda=0.75$ (ALMA Band 8), 0.89 (Band 7) and 1.34 mm (Band 6) computed from our models with the angular resolution of 0\farcs02.
We can see that the higher spatial resolution observations at shorter wavelengths can also distinguish between the active and passive disk models. 
The higher angular resolution allows us to detect a steeper brightness temperature profile in the active disk model.
However, the difference in the brightness temperatures of the active and passive disks decreases with observing wavelength decreases because the shorter wavelength traces a more upper layer where the accretion heating is less effective. 
The difference in the brightness temperatures of the two passive disk models is maximized at $\lambda\sim1.34$ mm because $300~{\rm \mu m}$-sized grain has a maximum effective albedo at that wavelength.
In contrast, the active disk models with $a_{\rm max}=300~{\rm \mu m}$ and 10 cm have similar brightness temperature profile because the $300~{\rm \mu m}$ dust model has a higher dust temperature, which compensates the efficient intensity reduction by scattering.

\subsection{Spectral energy distribution}
\begin{figure}[ht]
\begin{center}
\includegraphics[scale=0.5]{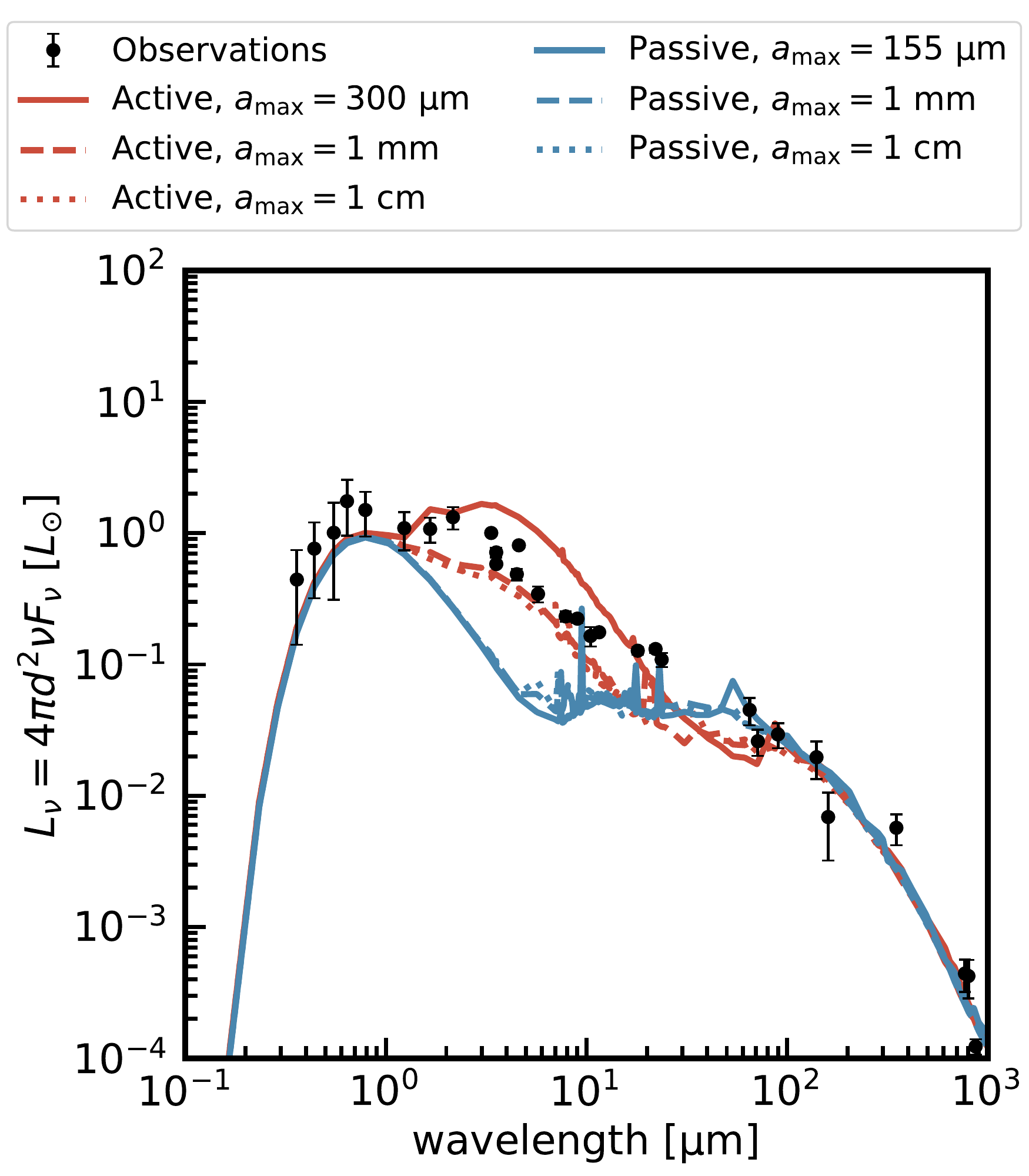}
\caption{
Comparison between our model spectral energy distributions and observed SED.
The observed values are from \citet{Andrews+13} and references therein.
}
\label{fig:SED}
\end{center}
\end{figure}

Although our focus is on comparing our models with observations at ALMA wavelengths, it is worthwhile to compare them at infrared wavelengths as well. 
Figure \ref{fig:SED} shows the comparison between our model spectral energy distributions (SEDs) and the observed data. 
The observed SED is taken from \citet{Andrews+13} (see also references therein).
In order to evaluate the SEDs, we extrapolate the disk model down to 0.01 au and remove the dust with temperature higher than 1400 K which mimics the sublimation of silicate dust.
This modification is necessary because the dust responsible for infrared emission could be located within 0.5 au, which is the inner boundary used for the calculations in the comparison at ALMA wavelengths.

We find that the active disk models have higher luminosity at infrared wavelengths ($\sim$1--20 ${\rm \mu m}$) than the passive disk models  (see also, e.g., \citealt{DAlessio+98,Dullemond+07}). 
The passive disk models tend to underestimate the fluxes observed at infrared wavelengths. 
For the active disk models, the large grain models ($a_{\rm max}=1$ mm and 1 cm) still underestimate the infrared fluxes, while the model with $a_{\rm max}=300~{\rm \mu m}$ overestimates them. 
Although neither of these models fully explains the observed SED at infrared wavelengths, the active disk models are more resemble to the observations compared to the passive disk models.

However, it is important to note that the infrared SED is highly influenced by the intricate structure of the innermost region of the disk, specifically in the vicinity of the silicate sublimation radius  (see \citealt{Dullemond+10} and \citealt{Kraus15} for review), which is beyond the scope of our study.
For instance, the emission in the near-infrared range ($\sim2~{\rm \mu m}$) is expected to be predominantly attributed to the scattering of stellar light at the inner rim of the disk (e.g., \citealt{MM02}), rather than the thermal emission from the region we focus on.
Additionally, the presence of a dust halo surrounding the inner disk, potentially induced by the inner disk wind, is anticipated to enhance the near-infrared flux (e.g., \citealt{VJ07}).
Therefore, detailed modeling of the innermost region of the disk is necessary to obtain  accurate fluxes at near-infrared wavelengths.
While significant uncertainties remain in the near-infrared range, the observed high fluxes at mid-infrared wavelengths may suggest a higher likelihood of active heating within the CW Tau disk.

\section{Summary} \label{sec:summary}

We investigate the impact of accretion heating on the brightness temperatures of the inner region of the CW Tau disk.
The key findings are as follows:

\begin{itemize}

\item If $a_{\rm max}\lesssim100~{\rm \mu m}$, the model brightness temperatures are too high to explain the observations regardless of the efficiency of accretion heating.

\item If the disk is passively heated, the maximum dust size needs to be either $\sim150~{\rm \mu m}$ or $\gtrsim$ few cm.

\item If the disk is actively heated, small grain models significantly overpredict the brightness temperatures, and hence
the maximum dust size needs to be either $\sim300~{\rm \mu m}$ or $\gtrsim$ few cm.

\item The midplane temperature would be $\sim$1.5--3 times higher than the observed brightness temperatures because of combined effect of scattering and accretion heating.

\item If dust settling occurs in the active disk, the ALMA observations trace deeper disk layers compared to the disk with no-settling. Therefore, the dust settling effectively increases the temperature of the dust responsible for the millimeter emission from the active disk.

\item If turbulence strength parameter for vertical dust mixing $\alpha_{\rm t}$ is $10^{-4}$, the active disk model with $a_{\rm max}=300~{\rm \mu m}$ overpredicts the brightness temperatures, suggesting that $\alpha_{\rm t}>10^{-4}$.

\item The estimated effective accretion efficiency $\alpha_{\rm acc}$ falls into a reasonable range of $10^{-4}$--$10^{-2}$, depending on the dust size. 
If $a_{\rm max}=300~{\rm \mu m}$, $\alpha_{\rm acc}$ is expected to be $\sim10^{-3}$, indicating that $\alpha_{\rm t}$ is not much smaller than $\alpha_{\rm acc}$.

\item The efficiency of accretion heating needs to be reduced if dust is porous ($p=0.9$) because the accretion heating is too efficient to explain the observations regardless of the dust size.

\end{itemize}
Future observations using longer wavelengths, such as ALMA Bands 1--3 and VLA, will play a crucial role in distinguishing between the active and passive disk models. The results of our work highlight the importance of multi-wavelength observations in improving our understanding of the physical properties of the inner regions of protoplanetary disks.

\begin{acknowledgements}
We thank the anonymous referee for useful comments.
We also thank Sean Andrews for providing the observed SED of CW Tau.
T.U. acknowledges the support of the DFG-Grant "Inside: inner regions of protoplanetary disks: simulations and observations" (FL 909/5-1).
S. O. is supported by JSPS KAKENHI Grant Numbers JP18H05438, JP20H00182, and JP20H01948.
M.F. acknowledges funding from the European Research Council (ERC) under the European Union’s Horizon 2020 research and innovation program (grant agreement No. 757957).
\end{acknowledgements}

%
%
\bibliographystyle{aa} 
\bibliography{reference} 

\begin{appendix}

\section{Midplane temperature of the passive disks}\label{sec:temp}
In this section, we examine the accuracy of our temperature of the passive disk models. In our temperature model (Equation \ref{eq:t_irr}), we do not consider the effect of scattering of the stellar irradiation and adopt constant value for the incident angle of the stellar irradiation $\psi$ (0.02).
To see the validity of our simplified model, we perform thermal Monte Carlo radiative transfer simulations with RADMC3D \citep{RADMC} with takeing the anisotropic scattering into account.

\begin{figure}[ht]
\begin{center}
\includegraphics[scale=0.50]{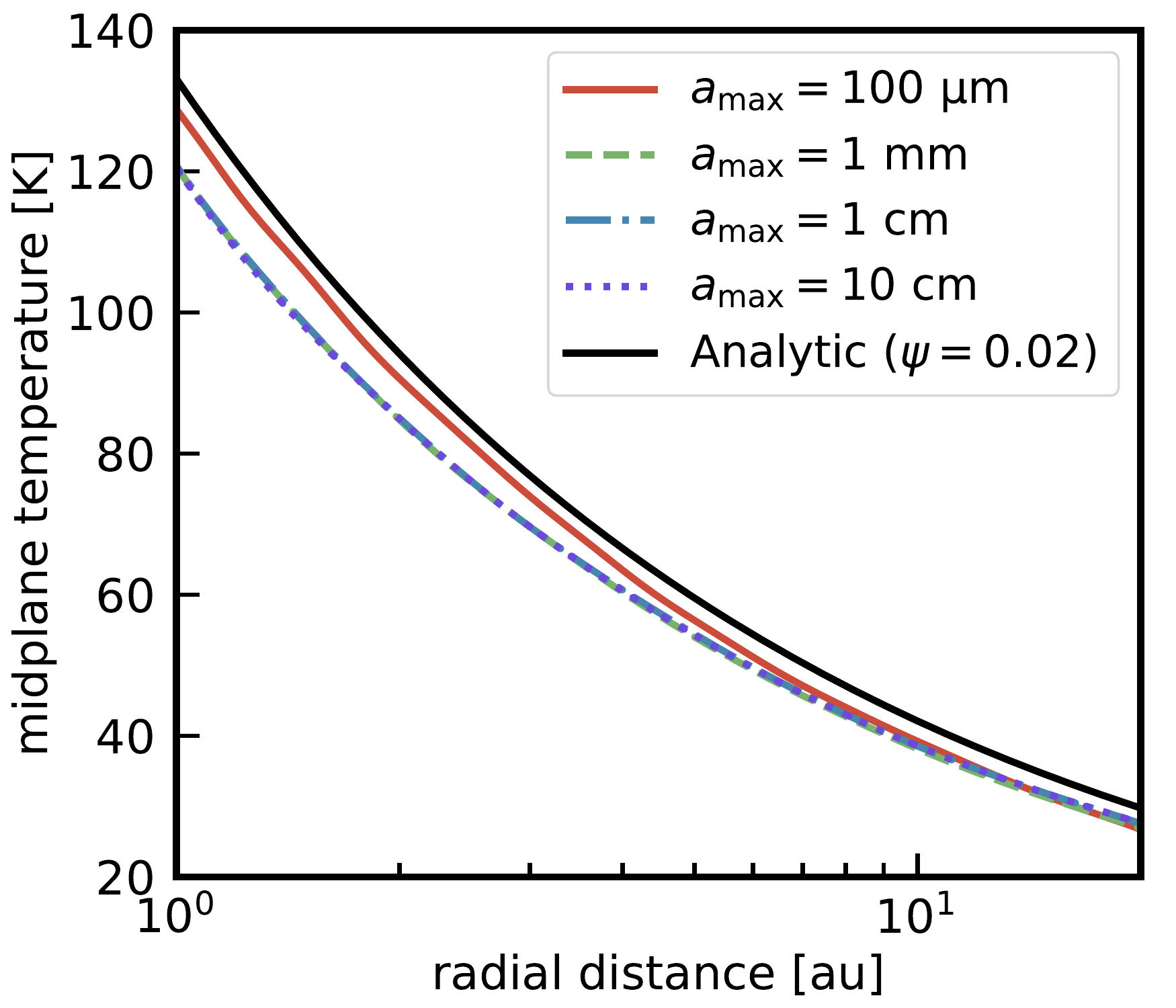}
\caption{
Comparison of the midplane temperature profiles of the passive disks obtained from the Monte Carlo radiative transfer simulations and analytical prescription (Equation \ref{eq:t_irr}).
}
\label{fig:temp_RADMC}
\end{center}
\end{figure}

Figure \ref{fig:temp_RADMC} compares the midplane temperature profiles obtained from the Monte Carlo radiative transfer simulations and analytical model (Equation \ref{eq:t_irr}).
Although our analytical model slightly overestimates the midplane temperature, the difference from the radiative transfer model is $\lesssim10$\% which is comparable to the uncertainty of the ALMA flux.
Therefore, we think that the analytical model has sufficient accuracy to estimate the brightness temperatures at the ALMA wavelengths.

\section{Effect of the dust composition} \label{sec:comp}
In this section, we present the modeling results using the DIANA dust, which is an alternative to the DSHARP dust used previously.
The DIANA dust consists of pyroxene (${\rm Mg_{0.7}Fe_{0.3}SiO_{3}}$; \citealt{Dorschner+95}) and amorphous carbon \citep{Zubko+96} , with a mass ratio of 0.87 for pyroxene and 0.13 for carbon, and has a porosity of 0.25 \citep{Woitke+16}.
While the DSHARP dust contains carbonaceous material in the form of organics, the DIANA dust consists of amorphous carbon as the carbonaceous material, which efficiently absorbs incident irradiation.

\begin{figure}[ht]
\begin{center}
\includegraphics[scale=0.43]{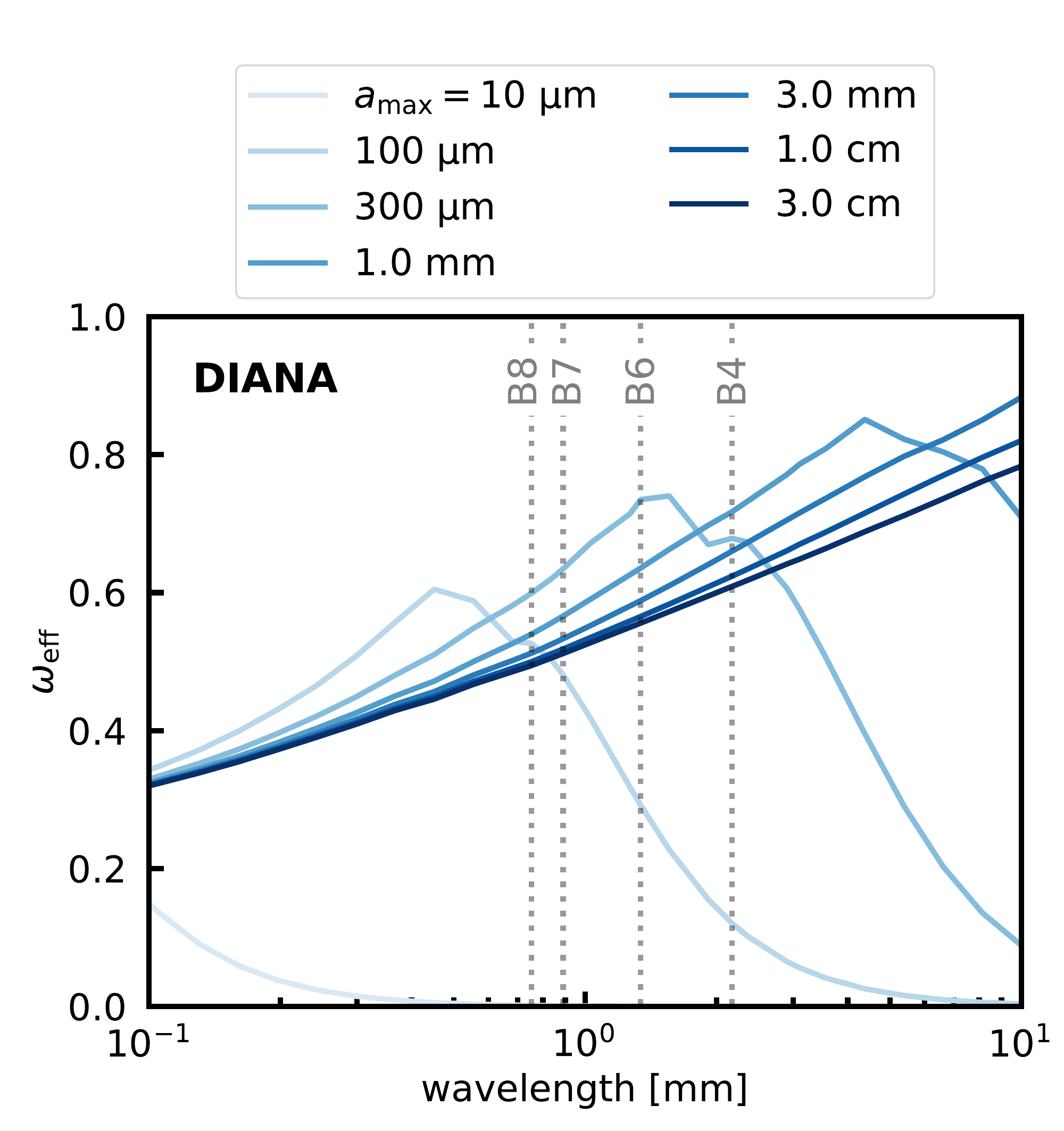}
\caption{
Effective albedo of the DIANA dust model.
}
\label{fig:opac_diana}
\end{center}
\end{figure}

Figure \ref{fig:opac_diana} shows the effective albedo of the DIANA dust.
The DIANA dust has lower effective albedo than the DSHARP dust because of the high absorbing efficiency of carbon.
Specifically, the effective albedo of the DIANA dust remains below 0.8 at the wavelengths of interest, while that of the DSHARP dust is as high as $\sim$ 0.9 (Figure \ref{fig:opac}).

\begin{figure}[ht]
\begin{center}
\includegraphics[scale=0.53]{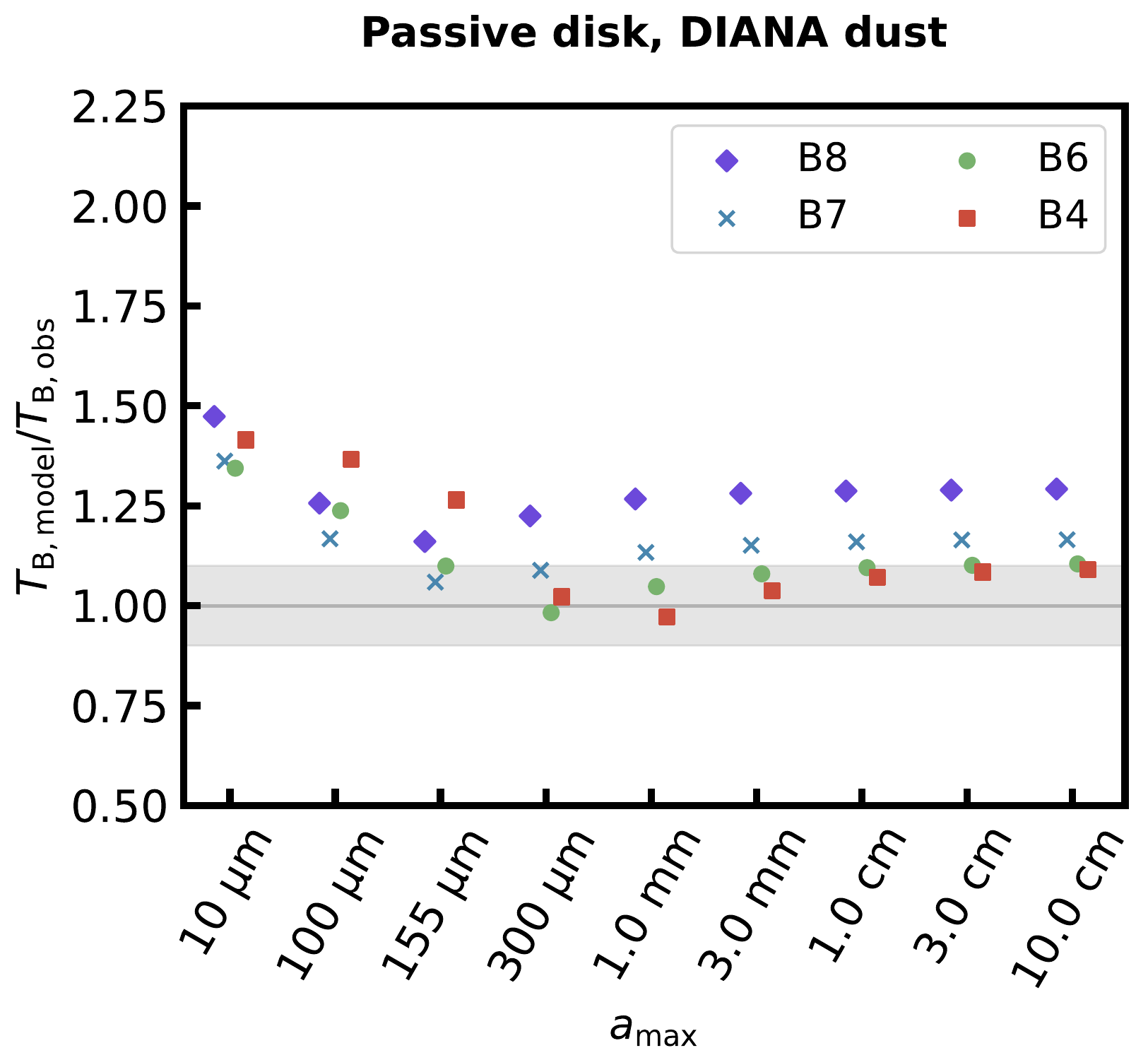}
\caption{
Same as Figure \ref{fig:passive_center} but for the DIANA dust model.
}
\label{fig:passive_diana}
\end{center}
\end{figure}

\begin{figure}[ht]
\begin{center}
\includegraphics[scale=0.53]{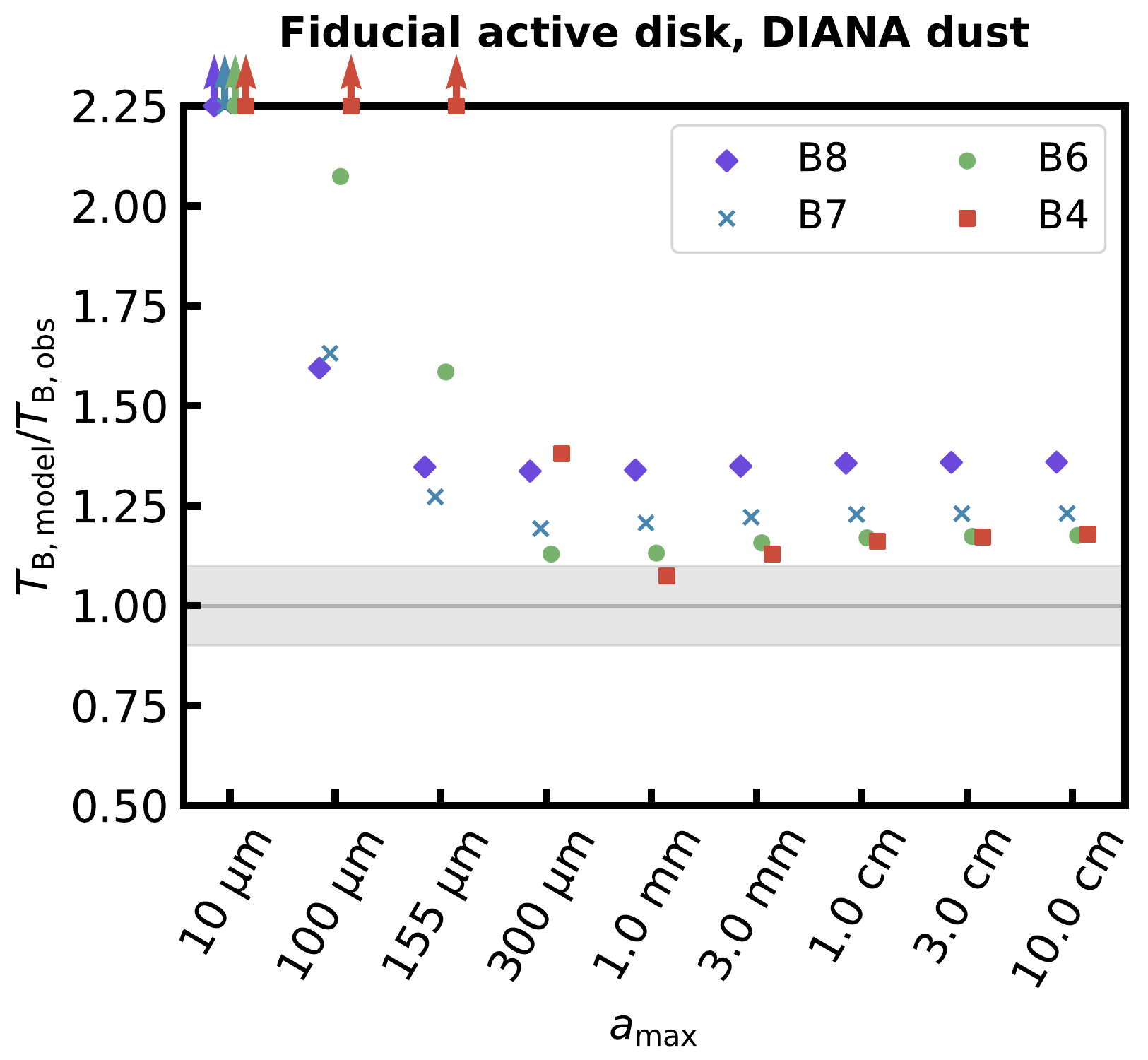}
\caption{
Same as Figure \ref{fig:active_center} but for the DIANA dust model.
}
\label{fig:active_diana}
\end{center}
\end{figure}

Figure \ref{fig:passive_diana} shows the normalized brightness temperature at the center of the passive disk model using the DIANA dust model.
Due to its smaller effective albedo, the brightness temperatures consistently exceed those of the DSHARP model (Figure \ref{fig:passive_center}).
Particularly at ALMA Band 8, the brightness temperature of the DIANA dust model is higher than the observed value, regardless of the calue of $a_{\rm max}$.
The central brightness temperature of the active disk model using the DIANA dust is shown in Figure \ref{fig:active_diana}.
When taking into account the effect of accretion heating, the brightness temperature is higher compared to the passive disk model, thereby making the DIANA dust model less consistent with the observations.

In general, the DIANA dust models predict higher brightness temperature than the DSHARP dust models. Consequently, the DSHARP dust model appears to be more in line with the observed data.

\section{Dependence of the surface density}\label{sec:app1}

\begin{figure*}[ht]
\begin{center}
\includegraphics[scale=0.53]{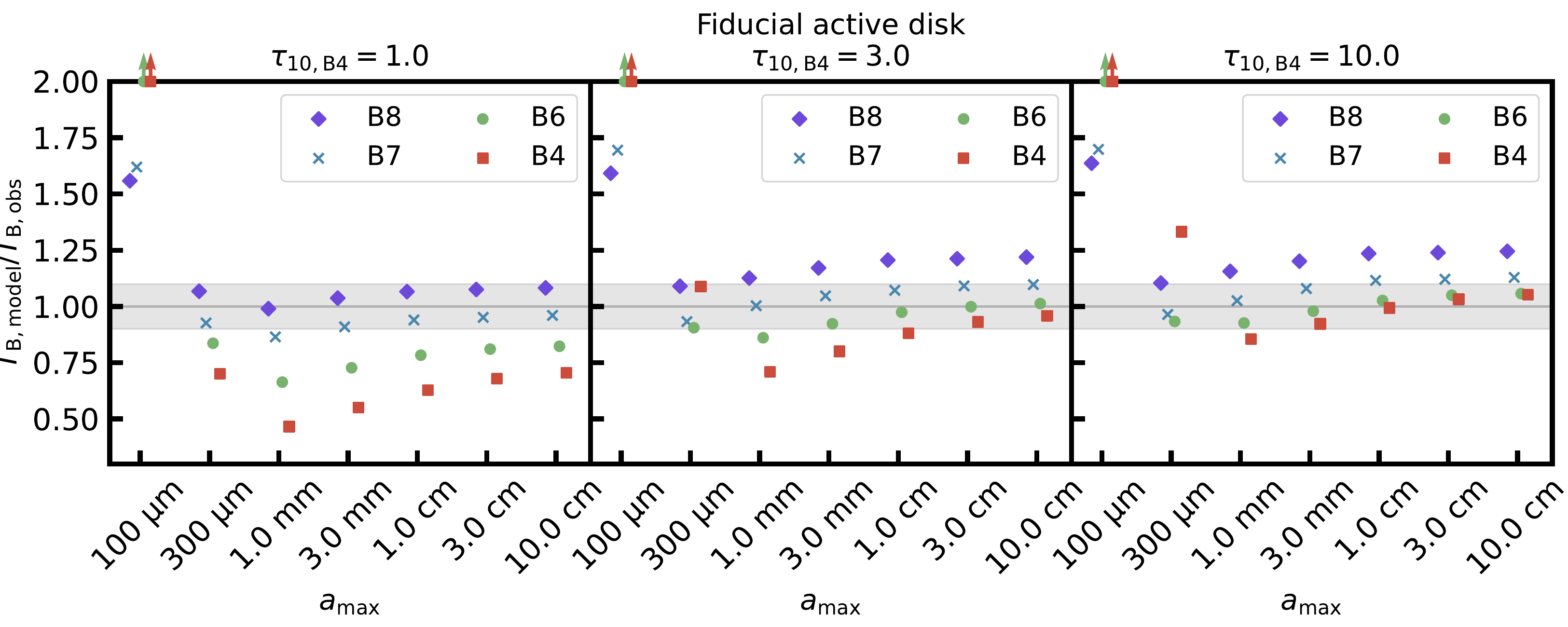}
\caption{
Same as Figure \ref{fig:active_center} but for different values of $\tau_{\rm 10,B4}$.
}
\label{fig:summary_app}
\end{center}
\end{figure*}
In this appendix, we discuss the effect of the dust surface density on the brightness temperature.
Figure \ref{fig:summary_app} shows the brightness temperature at the center of the synthetic images of different $\tau_{\rm 10,B4}$.
The brightness temperature at longer wavelength is more sensitive to the dust surface density (i.e., $\tau_{\rm 10,B4}$) because it traces the region closer to the midplane where the accretion heating dominates the irradiation heating.
In addition to that, smaller grain model is more sensitive to $\tau_{\rm 10,B4}$ because of efficient accretion heating (see Figure \ref{fig:opac-ratio}).

If $\tau_{\rm 10,B4}=1$, the brightness temperature at ALMA Band 4 is always lower than the observed value when $a_{\rm max}\gtrsim300~{\rm \mu m}$.
This suggests that $\tau_{\rm 10,B4}$ needs to be sufficiently larger than unity.
The brightness temperature at Band 4 increases as $\tau_{\rm 10,B4}$ increases.
If $\tau_{\rm 10,B4}=3$, the brightness temperature at Band 4 is consistent with the observation when $a_{\rm max}=300~{\rm \mu m}$, as shown in the main text.
If $\tau_{\rm 10,B4}>3$, the brightness temperature at Band 4 can be consistent with the observation even when $a_{\rm max}>300~{\rm \mu m}$.
However, in this regime, the brightness temperature at ALMA Band 8 is higher than the observed value.

\section{Radial brightness temperature profile of different models} \label{sec:app2}

\begin{figure*}[t]
\begin{center}
\includegraphics[scale=0.45]{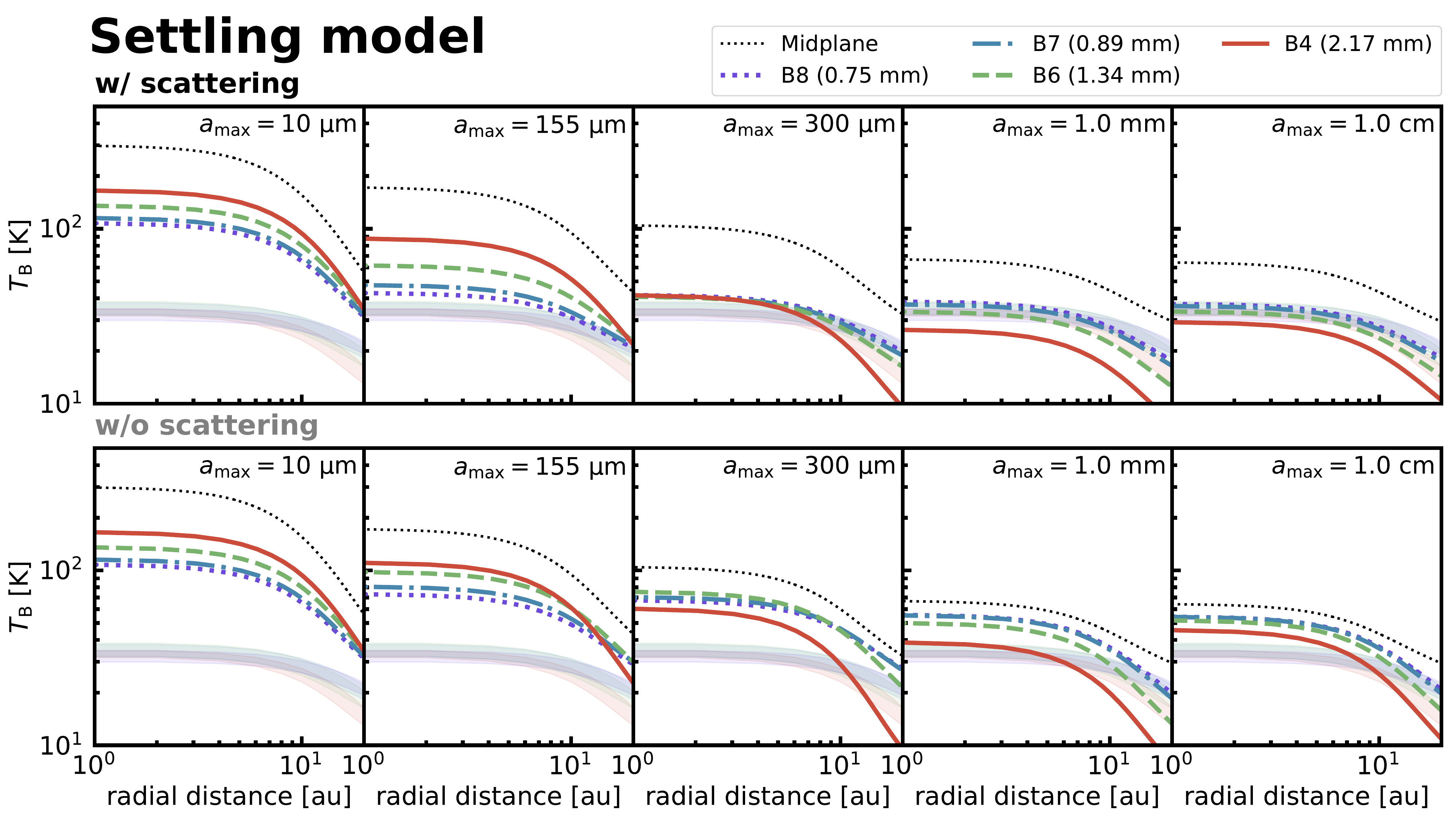}
\caption{
Same as Figure \ref{fig:1dtemp_fid} but with dust settling.
}
\label{fig:1dtemp_set}
\end{center}
\end{figure*}

Figure \ref{fig:1dtemp_set} shows the brightness temperature profile of models with dust settling.
The midplane temperature is independent on the presence of dust settling (see Figure \ref{fig:1dtemp_fid} and \ref{fig:1dtemp_set}) as the vertical optical depth at the midplane is identical.
The dust differential settling effectively increase the temperature of dust observed with ALMA as explained in Section \ref{sec:set}.
This can be confirmed by comparing the brightness temperature shown in Figure \ref{fig:1dtemp_fid} and \ref{fig:1dtemp_set} in the limit of no-scattering; The brightness temperature in the no-scattering limit is higher for the model with settling.
When scattering is taken into account, the brightness temperature of the settling model is not simply higher than the model without settling because the magnitude of scattering-induced intensity reduction is affected by the dust settling.

\begin{figure*}[t]
\begin{center}
\includegraphics[scale=0.45]{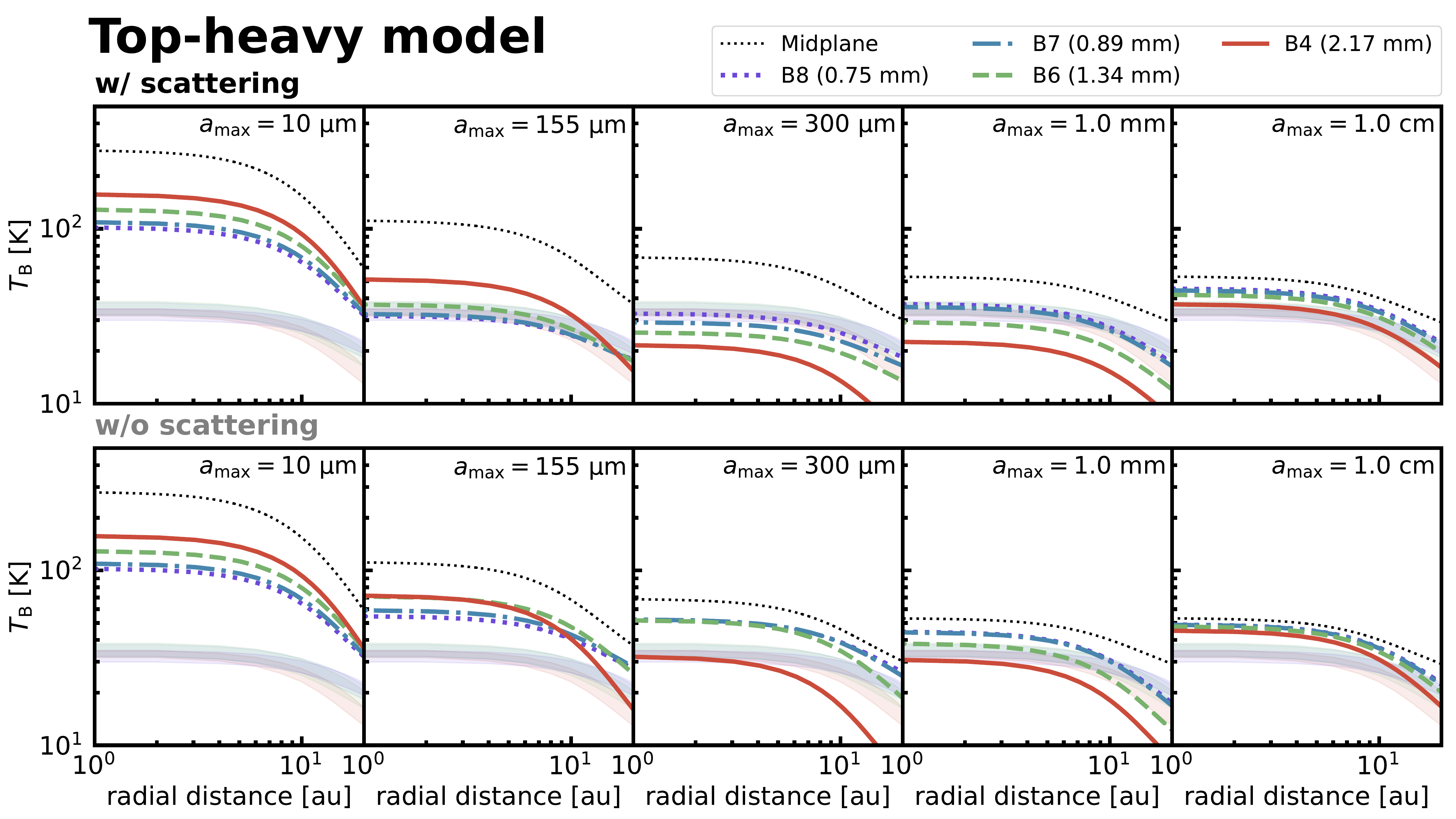}
\caption{
Same as Figure \ref{fig:1dtemp_fid} but with $p_{\rm d}=2.5$.
}
\label{fig:1dtemp_pd2.5}
\end{center}
\end{figure*}

In Figure \ref{fig:1dtemp_pd2.5}, we show the brightness temperature of models with $p_{\rm d}=2.5$ instead of 3.5.
The smaller $p_{\rm d}$ corresponds to the more top-heavy dust size distribution in which larger grains is more dominant in the opacity.
Because of the top-heavy dust size distribution, the dust temperature is lower than that of the fiducial active disk.
This can be confirmed by comparing the brightness temperature in Figure \ref{fig:1dtemp_fid} and \ref{fig:1dtemp_pd2.5} in the limit of no-scattering (dashed lines); The brightness temperature in the no-scattering limit is lower for the model with $p_{\rm d}=2.5$.
When scattering is taken into account, the brightness temperature of the top-heavy model is lower than that of the fiducial active disk in the regime of $a_{\rm max}\lesssim1~{\rm mm}$, because large grains have higher scattering albedo at ALMA wavelengths.
In contrast, in the regime of $a_{\rm max}\gtrsim1~{\rm mm}$, the brightness temperature of the top-heavy model is higher than the fiducial active disk model because of low scattering albedo.

\begin{figure*}[ht]
\begin{center}
\includegraphics[scale=0.45]{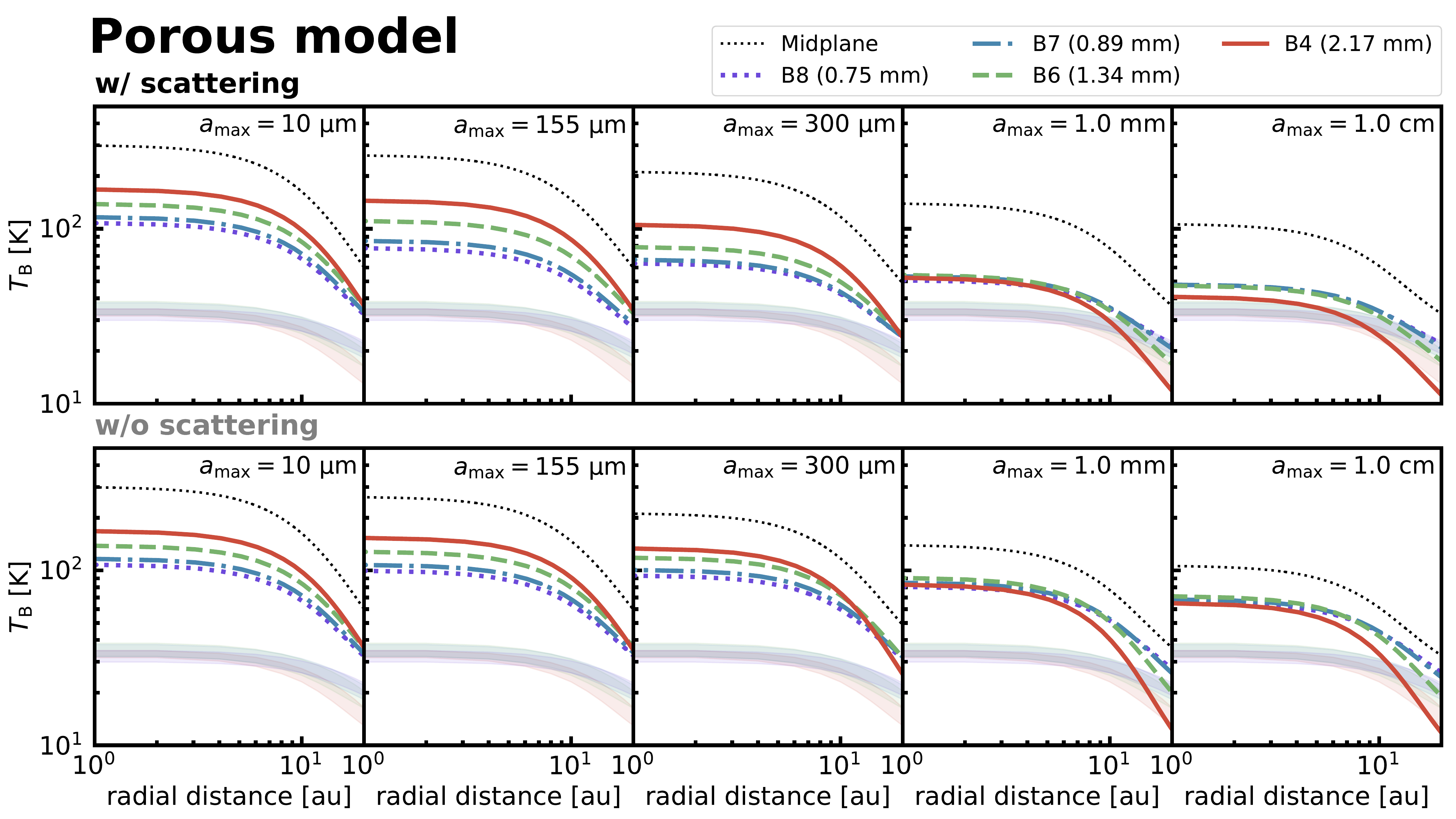}
\caption{
Same as Figure \ref{fig:1dtemp_fid} but with $p=0.9$.
}
\label{fig:1dtemp_porous}
\end{center}
\end{figure*}

Figure \ref{fig:1dtemp_porous} shows the brightness temperature profile of models with porous grains ($p=0.9$; filling factor of 0.1).

The dust temperature of the active disk with porous dust is higher than that of compact dust as explained in Section \ref{sec:por}.
This can be confirmed by comparing the brightness temperature in Figure \ref{fig:1dtemp_fid} and \ref{fig:1dtemp_porous} in the limit of no-scattering (dashed lines); The brightness temperature in the no-scattering limit is higher for the model with porous dust.
Furthermore, the effective scattering albedo of the porous dust is slightly lower than that of compact dust, which result in inefficient scattering-induced intensity reduction.
Because of these reasons, the model brightness temperature is significantly higher than the observations.

\end{appendix}

\end{document}